\documentclass[prl,twocolumn,nofootinbib,superscriptaddress]{revtex4}
\usepackage{amsmath,amssymb,epsfig,bm,mathrsfs,feynmp,slashed,color,graphicx,mathtools}
\usepackage[colorlinks=true,linkcolor=blue,citecolor=blue,urlcolor=blue]{hyperref}

\newcommand{\bea}{\begin{eqnarray}}
\newcommand{\eea}{\end{eqnarray}}
\newcommand{\ub}{\underbracket}

\newcommand{\De}{\Delta}
\newcommand{\ep}{\varepsilon}

\newcommand{\dyn}{{\rm dyn}}
\newcommand{\cm}{{\rm cm}}
\newcommand{\km}{{\rm km}}
\newcommand{\g}{{\rm g}}

\def\apj{ApJ \ }%
\def\apjs{ApJS}%
\def\aap{A\&A}%
\def\mnras{MNRAS}%
\def\prc{Phys. Rev. C\ }%
\def\prd{Phys. Rev. D \ }%
\def\sovast{Soviet Ast.}%
\def\nat{Nature}%
\def\XXint#1#2#3{{\setbox0=\hbox{$#1{#2#3}{\int}$}
     \vcenter{\hbox{$#2#3$}}\kern-.5\wd0}}

\definecolor{red}{rgb}{0.8,0,0}
\definecolor{violet}{rgb}{0.4,0,0.4}
\definecolor{green}{rgb}{0,0.5,0.0}
\definecolor{navy}{rgb}{0.0,0.0,0.6}
\definecolor{orange}{rgb}{0.8,0.2,0.0}

\usepackage[normalem]{ulem}  

\definecolor{red}{rgb}{0.8,0,0}
\definecolor{violet}{rgb}{0.4,0,0.4}
\definecolor{green}{rgb}{0,0.5,0.0}
\definecolor{navy}{rgb}{0.0,0.0,0.6}
\definecolor{orange}{rgb}{0.8,0.2,0.0}

\begin{document}
\title{
Compact Stars with Sequential QCD Phase Transitions 
}

\author{Mark Alford }
\affiliation{
Department of Physics, Washington University, St Louis, MO 63130, USA
}

\author{Armen Sedrakian} 
\affiliation{Frankfurt Institute for Advanced Studies, D-60438 
  Frankfurt-Main, Germany } 

\begin{abstract}
  Compact stars may contain quark matter in their interiors at
  densities exceeding several times the nuclear saturation density. We
  explore models of such compact stars where there are two first-order
  phase transitions: the first from nuclear matter to a quark-matter
  phase, followed at higher density by another first-order transition
  to a different quark matter phase [e.g., from the two-flavor color
  superconducting  (2SC) to the color-flavor-locked (CFL)
  phase).  We show that this can give rise to
  two separate branches of hybrid stars, separated from each other and
  from the nuclear branch by instability regions and, therefore,
  to a new family of compact stars, denser than the ordinary hybrid
  stars.  In a range of parameters, one may obtain twin hybrid stars
  (hybrid stars with the same masses but different radii) and even
  triplets where three stars, with inner cores of nuclear matter, 2SC
  matter, and CFL matter, respectively, all have the same mass but
  different radii.
\end{abstract}

\maketitle

\noindent {\it 1. Introduction.---}
Compact stars are formed in the last stages of stellar evolution,
their distinctive feature being that they are in gravitational
equilibrium supported by the quantum pressure of degenerate fermionic
matter.  The less dense of such objects, white dwarfs, are supported
by electron degeneracy pressure; the second densest class, neutron
stars, are supported by the degeneracy pressure of interacting
nucleonic (baryonic) matter.  It has been conjectured long ago
\cite{Ivanenko:1965,Pacini:1966,Boccaletti:1966,Itoh:1970uw} that a
higher-density class of compact stars may arise in the form of hybrid
(or quark) stars, whose core (or entire volume) consists of quark
matter.  It has been previously noted
\cite{1968PhRv..172.1325G,1977ApJ...213..840B,1981JPhA...14L.471K,2000A&A...353L...9G,Schertler:2000,2005PrPNP..54..193W}
that the hybrid stars may form a separate branch (third family) of
compact stars, separated from neutron stars by an instability region
analogous to the one existing between white dwarfs and neutron stars.
Such an elucidation of the relationship between the phases of
high-density matter and the observable properties of compact stars
helps to address one of the key challenges of strong interaction
physics, which is to constrain, from theory and experiment, the phase
diagram of ultradense matter.

NASA's NICER experiment, to become operative in 2017
\cite{2017AAS...22930903G}, will allow measurements of neutron star
masses and, especially, radii to unprecedented precision with better
than $10\%$ uncertainty.  Its capability of rotation-resolved
spectroscopy of the thermal and nonthermal emissions of neutron stars
in the soft ($0.2$--$12$\,keV) x-ray band is expected to provide new
insights into key properties of neutron stars, in particular
constraints on the mass-radius relation.  The measurements of the
radii of neutron stars in combination with the previously established
lower limit on the maximum mass of compact stars which is in the range
{$1.93(2)\,{M}_\odot$ \cite{2010Natur.467.1081D,Fonseca:2016tux}}
to $2.01(4)\,{M}_\odot$ \cite{2013Sci...340..448A} will strengthen
existing constraints on the equation of state (EOS) of ultradense
matter.

The main body of research on hybrid compact stars, i.e., stars that
are composed of a quark core surrounded by a nuclear envelope (which
in turn is composed minimally of a liquid core and a crust) has
concentrated on the case where the quark-matter core is represented by
a single phase (for recent reviews see
\cite{Brambilla:2014jmp,2014JPhG...41l3001B}).  However, as our
understanding of the QCD phase diagram improved over the years it
became clear that the quark core may contain layers of distinct phases
such as the various color superconducting phases of deconfined quark
matter \cite{2008RvMP...80.1455A,2014RvMP...86..509A}.  It is
generally agreed that the color-flavor-locked (CFL) phase will occur in the QCD phase
diagram at sufficiently high densities, but different quark-matter
phases may occur at intermediate densities, such as the two-flavor
color-superconducting (2SC) phase and
related phases
\cite{Ruester:2005jc,2005PhRvD..72f5020B,2009PhRvC..80f5807B,2012A&A...539A..16B},
unpaired quark matter \cite{Warringa:2006dk}, or other alternatives
\cite{Fukushima:2010bq}. The stability of the star sequences which
develop CFL matter cores has been questioned in studies based on the
Nambu--Jona-Lasinio
model~\cite{2004PhLB..595...36B,2007PhLB..654..170K,2013PhRvD..88h5001K},
but additional repulsive vector interaction appears to stabilize stars
with CFL cores~\cite{2012A&A...539A..16B}. Generically, repulsive
interaction in high-density (unpaired) quark matter leads also to
high-mass twin stars with and without strangeness degrees of freedom
~\cite{2015PhRvC..91e5808D,2016EPJA...52..232A,Alvarez-Castillo:2014dva,2016EPJA...52...69A,2017PhRvD..95j3008Z,2009PhRvD..79j3006A},
observations of which  may serve as 
  evidence of the existence of a critical end point in the QCD phase
  diagram~\cite{2015A&A...577A..40B,BlaPOS}.

  At densities near or below nuclear saturation density ($n_{\rm
    sat}=0.16$ fm$^{-3}$), we use a ``natural'' EOS, constructed from a
  Lagrangian (or Hamiltonian) that is fitted to nuclear phenomenology.
  At higher densities, we allow for two sharp first-order phase
    transitions, assuming that mixed phases are disfavored by surface
    tension and electrostatic energy costs
    \cite{Alford:2001zr,Palhares:2010be,Pinto:2012aq,Alvarez-Castillo:2014dva}.
    Since the phase structure in that region is unknown, we use a
    ``synthetic'' EOS via a ``CSS'' parameterization
    \cite{2013PhRvD..88h3013A} in which each quark-matter phase is
    assumed to have a constant (density-independent) speed of sound
    \cite{1971SvA....15..347S,2013A&A...551A..61Z,2013PhRvD..88h3013A}.
    This can describe the two sequential phase transitions in terms
  of six parameters (see below).  In this parameter space we explore
  the implications of such phase transitions for the masses and radii
  of compact stars.  We find that the second phase transition can lead
  to a new branch (fourth family) of compact stars, which in turn
  gives rise to new phenomena such as twin configurations where both
  members are hybrid stars and even triplets consisting of three
  distinct configurations with the same mass, but different radii and
  internal composition.

\begin{figure}[t] 
\includegraphics[width=0.9\hsize]{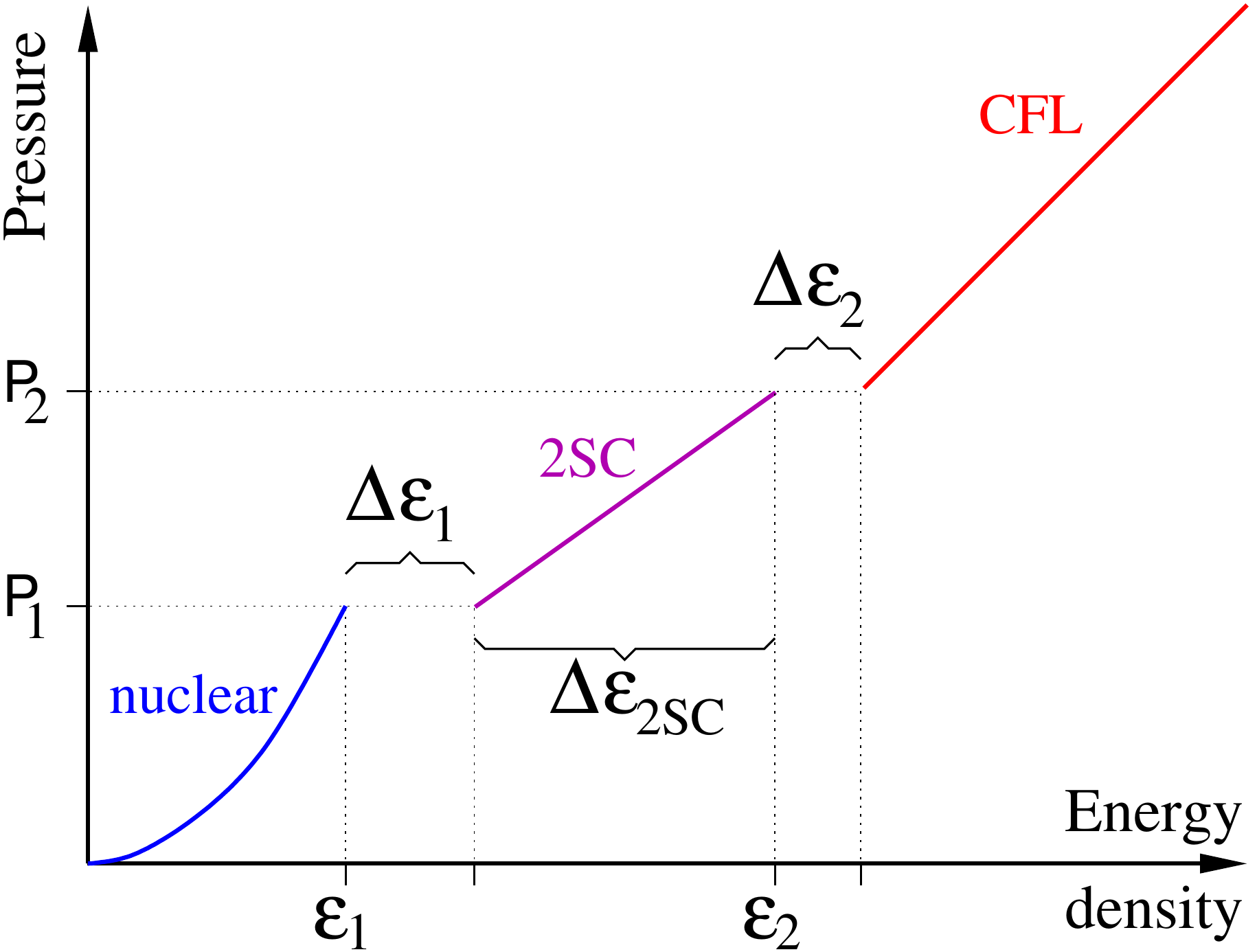}
\caption{ Schematic plot showing how we parametrize the EOS of dense
  matter with two phase transitions to two quark-matter phases.  For
  convenience and specificity, we call the first quark-matter phase
  2SC and the second CFL.  }
\label{fig:EOS_schematic}
\end{figure}

\medskip
\noindent {\it 2. Generating synthetic equations of state.}---  The
parameters of our EOS are illustrated in Fig.~\ref{fig:EOS_schematic}.
For nuclear matter we use the ``DDME2'' EOS which is based on
density-dependent relativistic functional theory
\cite{2013PhRvC..87e5806C}. This EOS fulfills the constraints derived
from the heavy ion collisions and other terrestrial experiments, see
Fig.~12 of Ref.~\cite{Fortin:2016hny}. It produces nucleonic compact
stars with a maximum gravitational mass $M\simeq 2.3\,{M}_{\odot}$,
where ${M}_{\odot}$ is the solar mass.
The quark-matter EOS is parametrized by \cite{2013PhRvD..88h3013A}\\
(i) $P_1$ and $P_2$ (or, equivalently, $P_1$ and $\De\ep_{\rm 2SC}$),
the transition pressures for the nuclear $\to$ 2SC and 2SC $\to$ CFL
transitions;\\
(ii) $\De\ep_1$ and $\De\ep_2$, the magnitudes of the jumps in the 
energy density at these two phase transitions; and \\
(iii) the squared sound speeds $s_1$ and $s_2$ in the 2SC and CFL phases.
Causality requires $s_{1,2}\leqslant 1$.

The analytic form of the quark-matter
EOS is
\bea\label{eq:synth_eos}
P(\ep) =\left\{
\begin{array}{ll}
P_{1}, &  \ep_1 < \ep < \ep_1\!+\!\De\ep_1\\[0.5ex]
P_1 + s_1 \bigl[\ep-(\ep_1\!+\!\De\ep_1)\bigr], 
  & \ep_1\!+\!\De\ep_1 < \ep < \ep_2\\[0.5ex]
P_2, & \ep_2 <\ep < \ep_2\!+\!\De\ep_2 \\[0.5ex]
P_2+ s_2\bigl[\ep-(\ep_2\!+\!\De\ep_2)\bigr], & \ep > \ep_2\!+\!\De\ep_2 \ .
\end{array}
\right.
\eea

\medskip
\noindent {\it 3.  Mass-radius relations of compact stars. ---} 
We solved the general relativistic structure equations of compact
stars \cite{Tolman,OV} for our model EOS (\ref{eq:synth_eos}) for
spherically symmetric (nonrotating and nonmagnetized) stars.  We look
for stable configurations using the Bardeen-Thorne-Meltzer criterion
\cite{BTM_methods}, which in our context states that a star is stable
if the mass is rising with the central pressure.  There may be other
nonradial instabilities but we leave a study of these to future
work.

\begin{figure}[t] 
\includegraphics[width=\hsize]{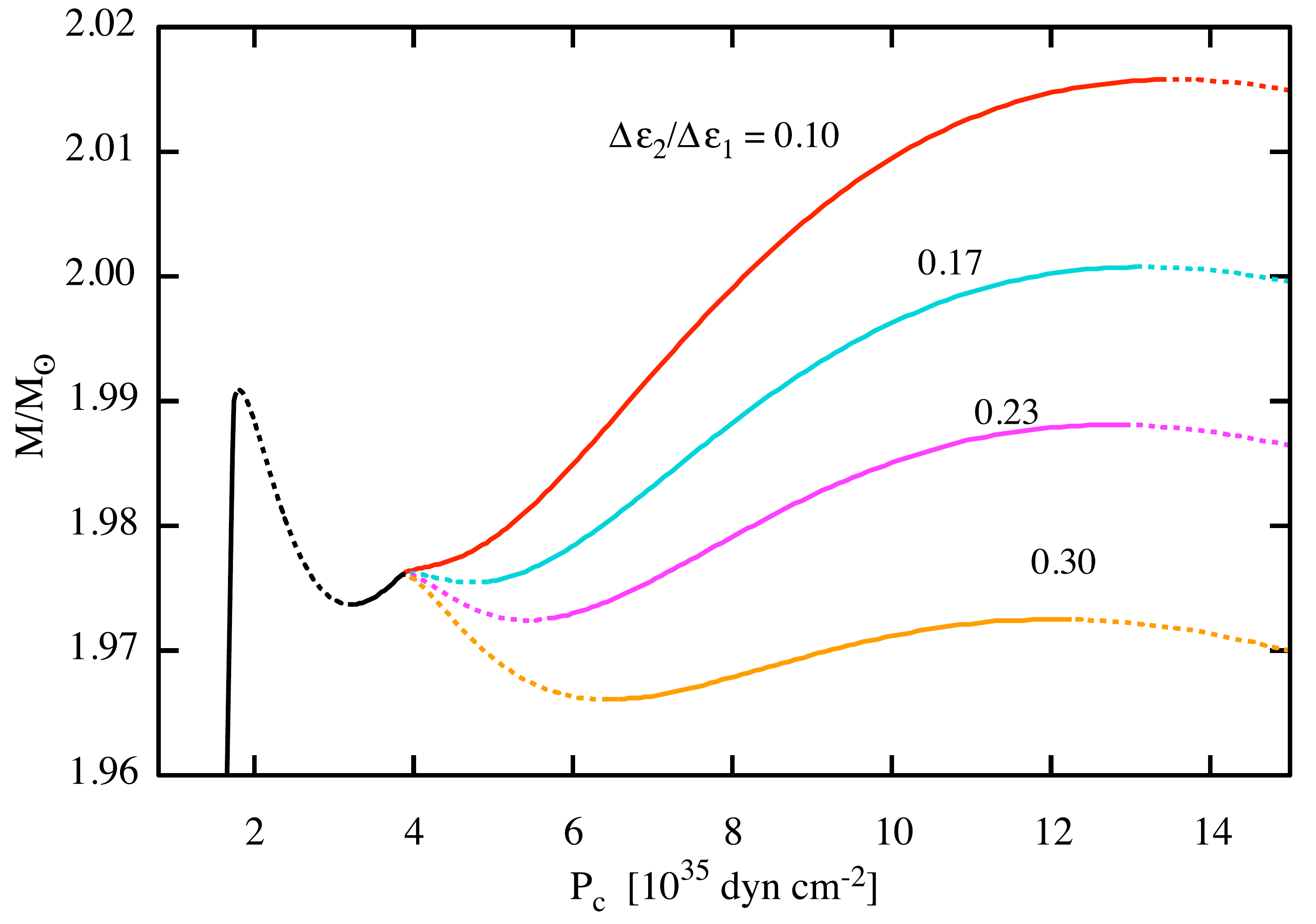}
\caption{The stellar mass as a function of the star's central
  pressure for four different values of $\De\ep_2$.  The other
  parameters of the EOS are fixed at
  $P_1=1.7 \times 10^{35}\,\dyn\,\cm^{-2}$,  $s_1 =0.7$,
  $\De\ep_{\rm 2SC}/\ep_1=0.27$, $\De\ep_1/\ep_1=0.6$, and $s_2=1$.  The
  vertical dotted lines mark the two phase transitions at $P_1$ and
  $P_2$.  Stable branches are solid lines, unstable branches are
  dashed lines.  We see the emergence of separate 2SC and CFL hybrid
  branches along with the occurrence of triplets.  }
\label{fig:M_vs_Pc}
\end{figure}

We first explore a scenario where both the quark-matter equations of
state are fairly stiff, with $s_1=0.7$ and $s_2=1$; we will discuss a
softer EOS for the 2SC phase below.  We fix the nuclear $\to$ 2SC
transition at $P_1=1.7\times 10^{35}\,\dyn\,\cm^{-2}$, corresponding
to nucleonic energy density $\ep_1=8.34\times 10^{14}\,\g\,\cm^{-3}$
and baryon density $n_1= 3.0n_{\rm sat}$.  This means that the mass
of the star reaches $M=1.99\,{M}_\odot$ before any transition to
quark matter occurs, ensuring that all our mass-radius curves obey the
observational lower bound on the maximum mass of a neutron star which
is in the range {$1.93(2)\,{ M}_\odot$ 
\cite{2010Natur.467.1081D,Fonseca:2016tux}} to
$2.01(4)\,{M}_\odot$ \cite{2013Sci...340..448A}.  There remain
three parameters to fix: the width of the 2SC phase $\De\ep_{\rm 2SC}$
and the two energy-density jumps $\De\ep_1$ and $\De\ep_2$.

Figure~\ref{fig:M_vs_Pc} shows the mass as a function of the central
pressure for four sequences of stars parametrized as follows.
We have fixed the width of the 2SC phase $\De\ep_{\rm 2SC}/\ep_1=0.27$
and the energy-density jump at the nuclear $\to$ 2SC transition
$\De\ep_1/\ep_1= 0.6$, and the four sequences have different values of
the energy-density jump $\De\ep_2$ at the 2SC $\to$ CFL transition.

Our choice of values of $P_1$, $\De\ep_1$, and $s_1$ leads to the
occurrence of a disconnected branch of stars with 2SC cores.  In the
figure, we see that, when the central pressure rises above $P_1$, and
2SC quark matter appears in the core, the star becomes unstable
(dashed black line), but then at
$P_c=3.2\times 10^{35}\,\dyn\,\cm^{-2}$ the stable branch of 2SC
hybrid stars begins (solid black line).

This sequence is then interrupted by the 2SC $\to$ CFL phase
transition at $P_2=3.11\times 10^{35}\,\dyn\,\cm^{-2}$, at which a CFL
core appears at the center of the star, within the existing 2SC core.
If there is a small energy-density jump $\De\ep_2$ at this transition
the hybrid branch will continue (upper solid line). However, if
$\De\ep_2$ is large enough, then the appearance of the CFL core
destabilizes the star again, until at a higher central pressure, thanks
to the stiffness of the CFL phase ($s_2=1$), a new stable sequence
emerges.

We see that for $\De\ep_2/\De\ep_1$ greater than about 0.15 
there are two separate, disconnected hybrid branches, both of which
are disconnected from the nucleonic branch of stars with $P_c<P_1$.
The disconnected stable branch of stars with a CFL core constitutes
a {\it fourth family} of compact stars, adding to
white dwarfs (not shown), ordinary neutron stars (near-vertical
black line at  $P_c<P_1$) and 2SC hybrid stars (solid black line
at $P_c$ just below $P_2$).

Moreover, for certain values of $\De\ep_2$ there exist {\it triplet
  configurations }: a set of three stars which have the same mass but
different central pressures, compositions, and radii.  In
Fig.~\ref{fig:M_vs_Pc} this is particularly clear for
$\De\ep_2/\De\ep_1=0.23$.

\begin{figure}[hbt] 
\includegraphics[width=\hsize]{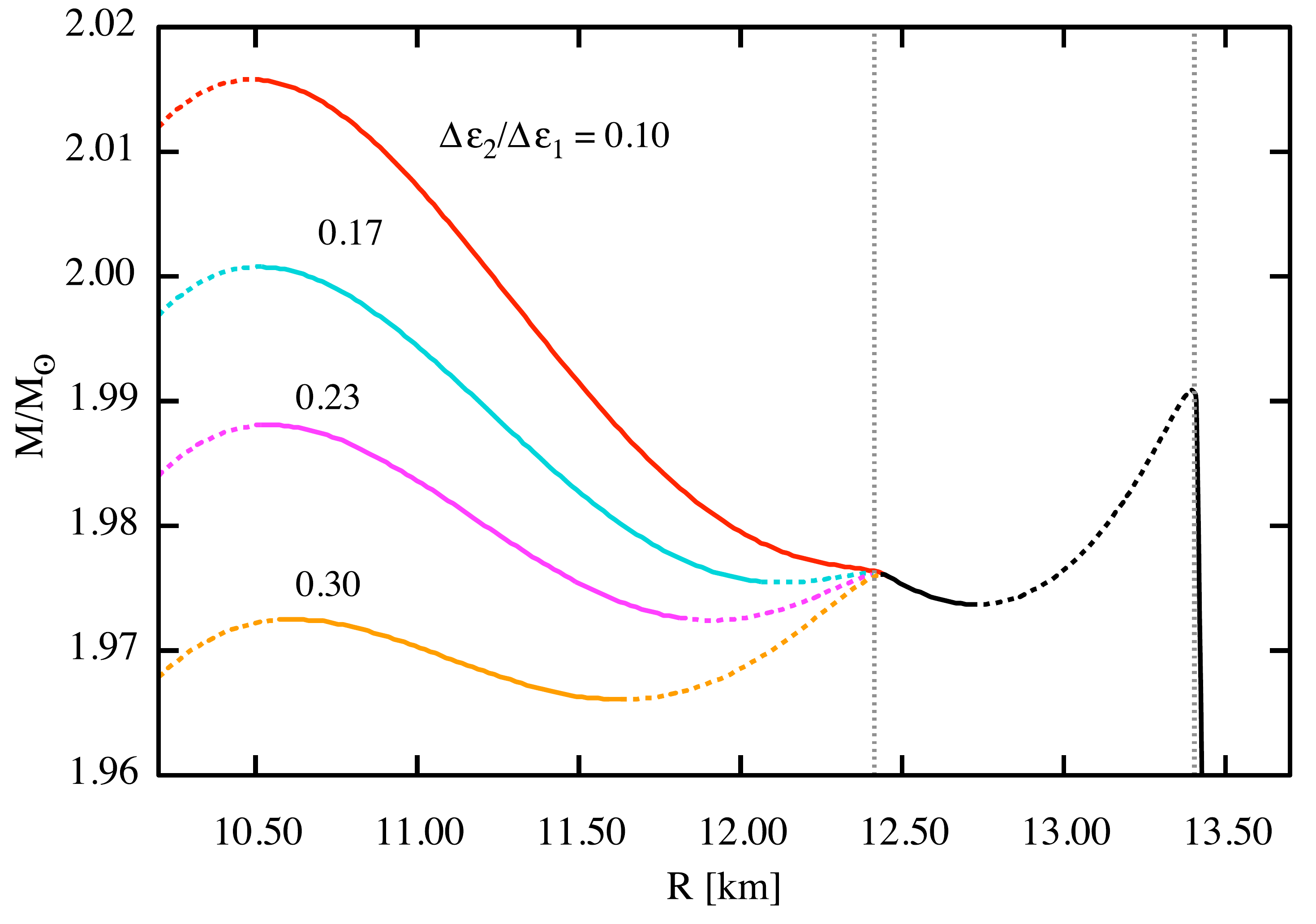}
  \caption{The $M$-$R$ relations for the parameter values defined in
    Fig.~\ref{fig:M_vs_Pc}. We have fixed the properties of the
nuclear $\to$ 2SC transition and the speed of sound in 2SC and CFL matter.
For the 2SC $\to$ CFL transition we have fixed the critical pressure
and we vary the energy-density discontinuity $\De\ep_2$. The separate
2SC and CFL hybrid branches are clearly visible, along with the
occurrence of triplets.}
\label{fig:M_vs_R_stiff2SC} 
\end{figure}

In Fig.~\ref{fig:M_vs_R_stiff2SC} we shown the mass-radius relation
for the sequences shown in Fig.~\ref{fig:M_vs_Pc}.  The disconnected
branches are, in principle, observable because they are separated by
intervals of radius in which no star can exist.  These disallowed
intervals cover ranges of hundreds of meters, which is only slightly
beyond the resolution of the measurements expected imminently from
NICER, and would provide motivation for future efforts to make radius
measurements more precise and increase the statistics. The observation
of two stars with very different radii will be a hint of the presence
of twins or triplets. In Fig.~\ref{fig:M_vs_R_stiff2SC}, the maximum
separation between the nucleonic branch and the CFL branch is about
$2$ km, well within NICER's resolution.

\begin{figure}[t] 
\includegraphics[width=\hsize]{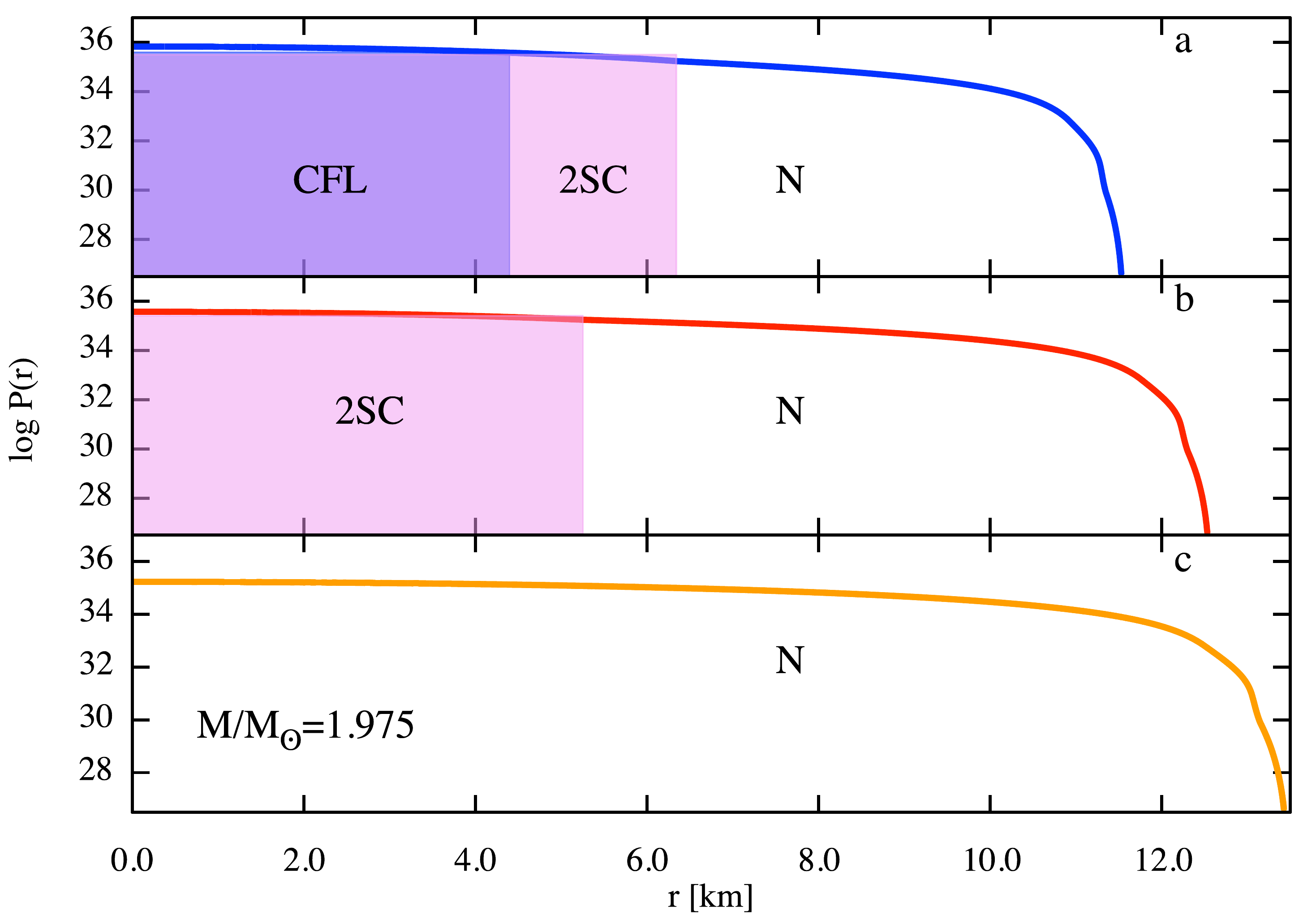}
\caption{The profiles (here the log of pressure as a function of the
  internal radius) of the three members of a triplet with masses
  $M=1.975\ {\rm M}_{\odot}$. Here ``$N$'' means the nuclear phase.
  The parameter values are the same as in Fig.~\ref{fig:M_vs_Pc}, with
  $\De\ep_2/\De\ep_1=0.23$.  }
\label{fig:profiles}
\end{figure}

In Fig.~\ref{fig:profiles}, we show the profiles of the three members
of the triplet of stars, all with mass $1.975M_\odot$, that occur
for the EOS parameter values used in Fig.~\ref{fig:M_vs_Pc}, with
$\De\ep_2/\De\ep_1=0.23$.  {The most compact member has a CFL core,
  and 2SC shell, with $R=11.5\,\km$.  The next has a 2SC core and
  $R=12.5\,\km$. The purely nucleonic member has $R=13.5\,\km$. The
  radii differ by 1--2 km, which is potentially detectable by
  NICER.}

The results shown in Figs.~\ref{fig:M_vs_Pc} and
\ref{fig:M_vs_R_stiff2SC} were for various values of $\De\ep_2$ at
fixed $\De\ep_1/\ep_1=0.6$. We now explore the effects of varying both
$\De\ep_1$ and $\De\ep_2$. Our results are summarized in
Table~\ref{table:1}. The notation describes the sequence of branches
encountered as the central pressure rises from $P_1$ up through $P_2$;
stable branches are denoted by ``$s$'' and unstable branches by ``$u$.''
A comma separates the 2SC sequence from the CFL sequence.  For
example, the top curve in Fig.~\ref{fig:M_vs_Pc} would be denoted
$us,s$ (unstable 2SC branch, stable 2SC branch, then a stable CFL
branch).  The bottom curve in Fig.~\ref{fig:M_vs_Pc} would be denoted
$us,us$ (unstable 2SC branch, stable 2SC branch, and then an unstable CFL
branch, and then a stable CFL branch).  Of course, all sequences
eventually become unstable at a high enough central pressure: We take
this as given and do not append a $u$ to every denotation.

\begin{table}
\begin{tabular}{c@{\quad}ccccc}
\hline
& \multicolumn{4}{c}{$\De\ep_1/\ep_1$} \\
\cline{2-5}
$\De\ep_2/\De\ep_1$  & 0.4   &    0.5 &    0.6 &    0.7
\\
\hline
0.1 &$s,s$&$s,s$                              &$\ub{us,s}_{\text {N-2SC}}$    & $\ub{u,us}_{\text{N-CFL}}$\\
0.2 &$s,s$&$s,s$                              &$\ub{us,us}_{\text{triplet}}$& $\ub{u,us}_{\text{N-CFL}}$\\
0.3 &$s,s$&$s,s$                              &$\ub{us,us}_{\text {N-2SC;N-CFL}}$& $\ub{u,us}_{\text{N-CFL}}$\\
0.4 &$s,s$&$\ub{s,us}_{\text{2SC-CFL}}$&$\ub{us,u}_{\text {N-2SC}}$& ${u,u}$\\
0.5 &$s,s$&$\ub{s,us}_{\text{2SC-CFL}}$&$\ub{us,u}_{\text {N-2SC}}$& ${u,u}$\\
\hline
\end{tabular}
\caption{
  Summary of the stability properties of
  compact star sequences as we vary the energy density 
  discontinuities $\De\ep_1$
  and $\De\ep_2$. {See the text for an explanation of the notation.}
The presence of twin hybrid configurations or triplet configurations
  is marked by the square underbraces
  with information about the involved phases (``N'' means nuclear).
  The fixed parameters $P_1$, $P_2$, $s_1$, and $s_2$
  are as in Figs.~\ref{fig:M_vs_Pc}
  and \ref{fig:M_vs_R_stiff2SC}.
}
\label{table:1}
\end{table}

When both the phase transition are weakly first order, with small
energy-density jumps (top left corner of Table~\ref{table:1}), the
phase transitions do not induce instabilities
\cite{1971SvA....15..347S,2013PhRvD..88h3013A}, so as the central
pressure rises above $P_1$ there is a single continuous family of
hybrid stars, denoted $s,s$, first with a 2SC core, and then with a
CFL core inside that at the center enveloped by a 2SC shell.

When both the phase transitions are strongly first order (bottom right
corner of Table~\ref{table:1}), the appearance of the denser phase
tends to destabilize the star, and both the 2SC and CFL sequences are
unstable (denoted as $u,u$).

From Table~\ref{table:1}, we see that the interesting phenomena
illustrated in Figs.~\ref{fig:M_vs_Pc} and \ref{fig:M_vs_R_stiff2SC}
arise as we vary the parameters of the EOS from the ``$s,s$'' domain
(no unstable branches) to the ``$u,u$'' domain (no stable branches).
In the intermediate parameter range, stability may be lost and regained
twice, once within the 2SC sequence and once within the CFL sequence,
creating the possibility of twin stars (with the same mass but
different radius) or even triplets (three stars with the same mass but
different radii).  Three types of twins are possible: N-2SC (hybrid
star with a 2SC core has the same mass as a nucleonic star), N-CFL
(hybrid star with a 2SC outer core and a CFL inner core has the same mass
as a nucleonic star), and 2SC-CFL (hybrid star with a 2SC core has the same
mass as a hybrid star with a 2SC outer core and  a CFL inner core). 

\begin{figure}[hbt] 
\includegraphics[width=\hsize]{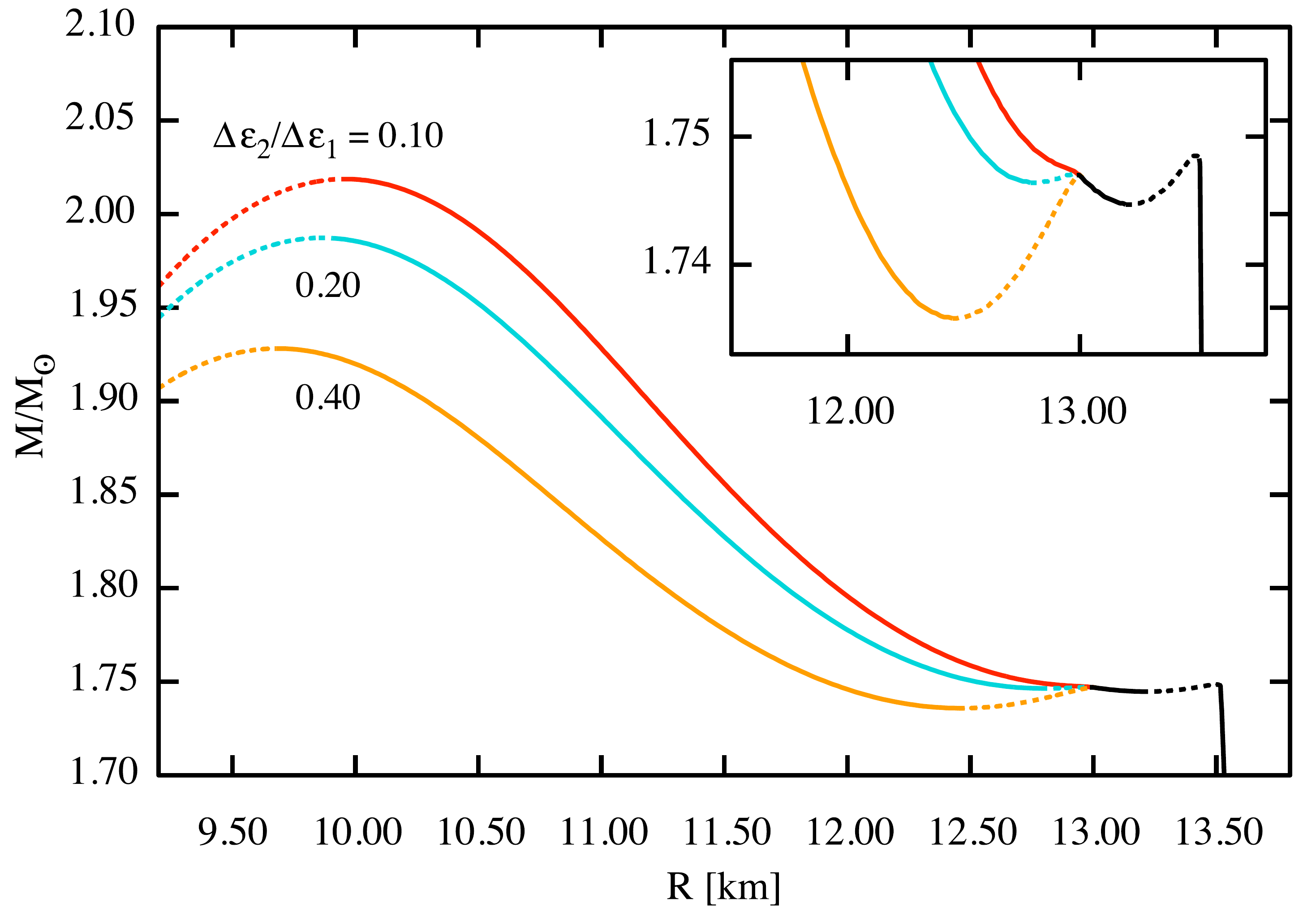}
\caption{The $M$-$R$ relation for a less stiff 2SC phase {($s_1=0.5$)}
  with four different values of $\De\ep_2$, keeping
  $\De\ep_1/\ep_1=0.6$.  The 2SC branch is shorter, but there can
  still be separate 2SC and CFL hybrid branches and triplets (the
  corresponding region is magnified in the figure inset).}
\label{fig:M_vs_R_soft2SC}
\end{figure}

The results above were obtained for stiff quark matter, with $s_1=0.7$
and $s_2=1$.  We now explore how our results change if the 2SC phase
is assumed to have a less stiff EOS with $s_1=0.5$, which is still
somewhat above the expected value $s=1/3$ for noninteracting massless
quarks.  The CFL phase remains maximally stiff, with $s_2=1$.  We
lower the nuclear $\to$ 2SC transition pressure to
$P_1=1.14\times 10^{35}\,\dyn\,\cm^{-2}$, corresponding to nucleonic
energy density $\ep_1=7.28\times 10^{14}\,\g\,\cm^{-3}$ and baryon
density $n_1= 2.6n_{\rm sat}$.  We set
$\De\ep_{\rm 2SC}/\ep_1=0.15$, corresponding to
$P_2=1.83\times 10^{35}\,\dyn\,\cm^{-2}$, and fix
$\De\ep_1/\ep_1=0.6$.

In Fig.~\ref{fig:M_vs_R_soft2SC} we show a set of mass-radius curves
for these parameter values, varying the energy-density discontinuity
$\De\ep_2$ at the 2SC $\to$ CFL transition.  In this case the
nucleonic branch ends at a mass of about $1.74\,M_\odot$, but
there are still families of stars that meet the maximum mass
constraint.  We see that the 2SC branch is shorter and shallower, but
there can still be separate 2SC and CFL hybrid branches, and triplets,
with ``forbidden'' ranges of radii covering several hundred meters.

A number of interesting astrophysical scenarios involve twins and by
extension also triplets discussed above. One scenario involves a
spin-up (in a binary) or spin-down (in isolation) induced QCD phase
transition in a compact star which would be accompanied by a quick
change in the star's global properties. This could induce drastic
(depending on how large the energy-density jump is) changes in spin,
for example,
backbending~\cite{1997PhRvL..79.1603G,2006A&A...450..747Z,2017A&A...600A..39B},
or a release of large portions of gravitational binding energy in an
explosion or
collapse~\cite{2008A&A...479..515Z,2009MNRAS.396.2269D,2009MNRAS.392...52A}.
Core-collapse supernovas provide yet another setting where the QCD
phase transition(s) can induce additional shock
wave(s)~\cite{2011ApJS..194...39F} and affect the supernova
outcome. These require an extension of the input EOS to finite
temperatures; see, e.g., Ref.~\cite{2016PhRvD..94j3008H}. 
  Finally, future detections of gravitational waves from binary
  neutron star inspirals and mergers may provide independent
  constraints on the radii and masses of compact objects; any density
  discontinuities in the EOS are likely to leave their distinctive
  imprint in the data which would reveal a phase transition(s) to QCD
  matter~\cite{Sotani:2001bb,Marranghello:2002yx,Miniutti:2002bh,Oechslin:2004yj,Bauswein:2011tp,Takami:2014zpa}.

\medskip
\noindent {\it 4. Conclusions and Outlook.}---  In this work, we investigated
the physical consequences of assuming that there are two sequential
first-order phase transitions in dense matter, first from a
nucleonic phase to a quark-matter phase that for convenience we called
2SC and second from that phase to a denser quark-matter phase that
we called CFL. (Such sequential first-order phase transitions emerge,
for example, in QCD-inspired models of dense 
quark-matter~\cite{Ruester:2005jc,2005PhRvD..72f5020B,2009PhRvC..80f5807B,2012A&A...539A..16B}).
By using simple constant-sound-speed parameterizations of the 
quark-matter EOS we were able to explore, at least partly, the spaces of
possible EOS and the mass-radius properties of the resulting stellar
sequences.  The models were constrained to be causal
($s_{1,2}\leqslant 1$) and to satisfy the two-solar-mass observational
constraint.

We found that if the quark matter is fairly stiff (the squared speed
of sound being at least $0.5$ in the 2SC phase, and 1 in CFL), then the
sequence of two phase transitions {can yield} characteristic
phenomena:\\
(a) {Pairs} of disconnected branches of hybrid stars, separated from
each other and from the nucleonic stars by unstable intervals,
corresponding to ranges of radii in which no stars can occur.  This
represents a new branch of compact stars (fourth family) which, for a
given EOS, are denser than the hybrid stars that arise when there is a
single phase transition from nucleonic to quark matter.  (b) Connected
with this, we find equal mass ``twin'' stars of the type N-2SC,
N-CFL which could have been anticipated from the studies of ordinary
hybrid stars, but also 2SC$-$CFL twins which both contain quark
matter.  (c) Triplet configurations: three equal-mass stars with
different radii and internal structures.

In the future, it would be useful to perform a more comprehensive
survey of the six-dimensional parameter space of our model, looking
for regularities and systematic features, and to match the
parameterization used in this study with models based on different
classes of QCD models (ranging from the perturbative to QCD-inspired
effective ones; see Refs. \cite{Brambilla:2014jmp,2014JPhG...41l3001B},
and references therein). Extrapolating from the current model, each
additional first-order phase transitions may lead to another
disconnected branch of compact stars.  Imminent observational
advances, in particular, the science program of the NICER experiment,
are expected to provide further insight on the potentially complex
structure of the compact stars that can exist in nature.

\begin{acknowledgments}
  We thank D.~Blaschke, Th.~Kl\"ahn, M.~Oertel, S.~Reddy,
  S. Schramm, L. Tol\'os, F. Weber, and D. Zappal\`a for useful
  comments on the manuscript.  M. A. is partly supported by the
  U.S. Department of Energy, Office of Science, Office of Nuclear
  Physics under Award No. \#DE-FG02-05ER41375.  A.S. is supported
  by the Deutsche Forschungsgemeinschaft (Grant No. SE 1836/3-2) and
  by NewCompStar COST Action MP1304.
\end{acknowledgments}


\begin{thebibliography}{53}%
\makeatletter
\providecommand \@ifxundefined [1]{%
 \@ifx{#1\undefined}
}%
\providecommand \@ifnum [1]{%
 \ifnum #1\expandafter \@firstoftwo
 \else \expandafter \@secondoftwo
 \fi
}%
\providecommand \@ifx [1]{%
 \ifx #1\expandafter \@firstoftwo
 \else \expandafter \@secondoftwo
 \fi
}%
\providecommand \natexlab [1]{#1}%
\providecommand \enquote  [1]{``#1''}%
\providecommand \bibnamefont  [1]{#1}%
\providecommand \bibfnamefont [1]{#1}%
\providecommand \citenamefont [1]{#1}%
\providecommand \href@noop [0]{\@secondoftwo}%
\providecommand \href [0]{\begingroup \@sanitize@url \@href}%
\providecommand \@href[1]{\@@startlink{#1}\@@href}%
\providecommand \@@href[1]{\endgroup#1\@@endlink}%
\providecommand \@sanitize@url [0]{\catcode `\\12\catcode `\$12\catcode
  `\&12\catcode `\#12\catcode `\^12\catcode `\_12\catcode `\%12\relax}%
\providecommand \@@startlink[1]{}%
\providecommand \@@endlink[0]{}%
\providecommand \url  [0]{\begingroup\@sanitize@url \@url }%
\providecommand \@url [1]{\endgroup\@href {#1}{\urlprefix }}%
\providecommand \urlprefix  [0]{URL }%
\providecommand \Eprint [0]{\href }%
\providecommand \doibase [0]{http://dx.doi.org/}%
\providecommand \selectlanguage [0]{\@gobble}%
\providecommand \bibinfo  [0]{\@secondoftwo}%
\providecommand \bibfield  [0]{\@secondoftwo}%
\providecommand \translation [1]{[#1]}%
\providecommand \BibitemOpen [0]{}%
\providecommand \bibitemStop [0]{}%
\providecommand \bibitemNoStop [0]{.\EOS\space}%
\providecommand \EOS [0]{\spacefactor3000\relax}%
\providecommand \BibitemShut  [1]{\csname bibitem#1\endcsname}%
\let\auto@bib@innerbib\@empty
\bibitem [{\citenamefont {Ivanenko}\ and\ \citenamefont
  {Kurdgelaidze}(1965)}]{Ivanenko:1965}%
  \BibitemOpen
  \bibfield  {author} {\bibinfo {author} {\bibfnamefont {D.~D.}\ \bibnamefont
  {Ivanenko}}\ and\ \bibinfo {author} {\bibfnamefont {D.~F.}\ \bibnamefont
  {Kurdgelaidze}},\ }\href@noop {} \textcolor{blue}{ {\bibfield  {journal} {\bibinfo  {journal}
  {Astrophysics}\ }\textbf {\bibinfo {volume} {1}},\ \bibinfo {pages} {479}
  (\bibinfo {year} {1965})}}\BibitemShut {NoStop}%
\bibitem [{\citenamefont {Pacini}(1966)}]{Pacini:1966}%
  \BibitemOpen
  \bibfield  {author} {\bibinfo {author} {\bibfnamefont {F.}~\bibnamefont
  {Pacini}},\ }\href@noop {} \textcolor{blue}{ {\bibfield  {journal}{ \bibinfo  {journal}
  {Nature (London)}\ }\textbf {\bibinfo {volume} {209}},\ \bibinfo {pages} {389}
  (\bibinfo {year} {1966})}}\BibitemShut {NoStop}%
\bibitem [{\citenamefont {Boccaletti}\ \emph {et~al.}(1966)\citenamefont
  {Boccaletti}, \citenamefont {de~Sabbata},\ and\ \citenamefont
  {Gualdi}}]{Boccaletti:1966}%
  \BibitemOpen
  \bibfield  {author} {\bibinfo {author} {\bibfnamefont {D.}~\bibnamefont
  {Boccaletti}}, \bibinfo {author} {\bibfnamefont {V.}~\bibnamefont
  {de~Sabbata}}, \ and\ \bibinfo {author} {\bibfnamefont {C.}~\bibnamefont
  {Gualdi}},\ }\href@noop {} \textcolor{blue}{ {\bibfield  {journal} {\bibinfo  {journal} {Nuovo
  Cimento}\ }\textbf {\bibinfo {volume} {45}},\ \bibinfo {pages} {513} (\bibinfo
  {year} {1966})}}\BibitemShut {NoStop}%
\bibitem [{\citenamefont {Itoh}(1970)}]{Itoh:1970uw}%
  \BibitemOpen
  \bibfield  {author} {\bibinfo {author} {\bibfnamefont {N.}~\bibnamefont
  {Itoh}},\ }\href@noop {} \textcolor{blue}{ {\bibfield  {journal} {\bibinfo  {journal} {Prog.
  Theor. Phys.}\ }\textbf {\bibinfo {volume} {44}},\ \bibinfo {pages} {291}
  (\bibinfo {year} {1970})}}\BibitemShut {NoStop}%
\bibitem [{\citenamefont {{Gerlach}}(1968)}]{1968PhRv..172.1325G}%
  \BibitemOpen
  \bibfield  {author} {\bibinfo {author} {\bibfnamefont {U.~H.}\ \bibnamefont
  {{Gerlach}}},\ }\href {\doibase 10.1103/PhysRev.172.1325} {\bibfield
  {journal} {\bibinfo  {journal} {Phys. Rev.}\ }\textbf {\bibinfo {volume}
  {172}},\ \bibinfo {pages} {1325} (\bibinfo {year} {1968})}\BibitemShut
  {NoStop}%
\bibitem [{\citenamefont {{Bowers}}\ \emph {et~al.}(1977)\citenamefont
  {{Bowers}}, \citenamefont {{Gleeson}},\ and\ \citenamefont
  {{Pedigo}}}]{1977ApJ...213..840B}%
  \BibitemOpen
  \bibfield  {author} {\bibinfo {author} {\bibfnamefont {R.~L.}\ \bibnamefont
  {{Bowers}}}, \bibinfo {author} {\bibfnamefont {A.~M.}\ \bibnamefont
  {{Gleeson}}}, \ and\ \bibinfo {author} {\bibfnamefont {R.~D.}\ \bibnamefont
  {{Pedigo}}},\ }\href {\doibase 10.1086/155216} {\bibfield  {journal}
  {\bibinfo  {journal} {\apj}\ }\textbf {\bibinfo {volume} {213}},\ \bibinfo
  {pages} {840} (\bibinfo {year} {1977})}\BibitemShut {NoStop}%
\bibitem [{\citenamefont {{K\"ampfer}}(1981)}]{1981JPhA...14L.471K}%
  \BibitemOpen
  \bibfield  {author} {\bibinfo {author} {\bibfnamefont {B.}~\bibnamefont
  {{K\"ampfer}}},\ }\href {\doibase 10.1088/0305-4470/14/11/009} {\bibfield
  {journal} {\bibinfo  {journal} {J. of Phys. A}\ }\textbf {\bibinfo
  {volume} {14}},\ \bibinfo {pages} {L471} (\bibinfo {year}
  {1981})}\BibitemShut {NoStop}%
\bibitem [{\citenamefont {{Glendenning}}\ and\ \citenamefont
  {{Kettner}}(2000)}]{2000A&A...353L...9G}%
  \BibitemOpen
  \bibfield  {author} {\bibinfo {author} {\bibfnamefont {N.~K.}\ \bibnamefont
  {{Glendenning}}}\ and\ \bibinfo {author} {\bibfnamefont {C.}~\bibnamefont
  {{Kettner}}},\ }\href@noop {} {\bibfield  {journal} {\bibinfo  {journal}
  {\aap}\ }\textbf {\bibinfo {volume} {353}},\ \bibinfo {pages} {L9} (\bibinfo
  {year} {2000})},\ \Eprint {http://arxiv.org/abs/astro-ph/9807155}
  {astro-ph/9807155} \BibitemShut {NoStop}%
\bibitem [{\citenamefont {{Schertler}}\ \emph {et~al.}(2000)\citenamefont
  {{Schertler}}, \citenamefont {{Greiner}}, \citenamefont
  {{Schaffner-Bielich}},\ and\ \citenamefont {{Thoma}}}]{Schertler:2000}%
  \BibitemOpen
  \bibfield  {author} {\bibinfo {author} {\bibfnamefont {K.}~\bibnamefont
  {{Schertler}}}, \bibinfo {author} {\bibfnamefont {C.}~\bibnamefont
  {{Greiner}}}, \bibinfo {author} {\bibfnamefont {J.}~\bibnamefont
  {{Schaffner-Bielich}}}, \ and\ \bibinfo {author} {\bibfnamefont {M.~H.}\
  \bibnamefont {{Thoma}}},\ }\href {\doibase 10.1016/S0375-9474(00)00305-5}
  {\bibfield  {journal} {\bibinfo  {journal} {Nucl. Phys.}\ }\textbf {\bibinfo
  {volume} {A677}},\ \bibinfo {pages} {463} (\bibinfo {year} {2000})},\ \Eprint
  {http://arxiv.org/abs/astro-ph/0001467} {arXiv:astro-ph/0001467 [astro-ph]}
 \BibitemShut {NoStop}%
\bibitem [{\citenamefont {{Weber}}(2005)}]{2005PrPNP..54..193W}%
  \BibitemOpen
  \bibfield  {author} {\bibinfo {author} {\bibfnamefont {F.}~\bibnamefont
  {{Weber}}},\ }\href {\doibase 10.1016/j.ppnp.2004.07.001} {\bibfield
  {journal} {\bibinfo  {journal} {Prog. in Part.  Nucl. Phys.}\
  }\textbf {\bibinfo {volume} {54}},\ \bibinfo {pages} {193} (\bibinfo {year}
  {2005})},\ \Eprint {http://arxiv.org/abs/astro-ph/0407155} {astro-ph/0407155}
 \BibitemShut {NoStop}%
\bibitem [{\citenamefont {{Gendreau}}\ \emph {et~al.}(2017)\citenamefont
  {{Gendreau}}, \citenamefont {{Arzoumanian}},\ and\ \citenamefont {{NICER
  Team}}}]{2017AAS...22930903G}%
  \BibitemOpen
  \bibfield  {author} {\bibinfo {author} {\bibfnamefont {K.}~\bibnamefont
  {{Gendreau}}}, \bibinfo {author} {\bibfnamefont {Z.}~\bibnamefont
  {{Arzoumanian}}}, \ and\ \bibinfo {author} {\bibnamefont {{NICER
      Team, AAS Meeting}}}\
  }(\bibinfo {year} {2017})\ \bibinfo {journal} {No. 229, id. 309.03}\BibitemShut
  {NoStop}%
\bibitem [{\citenamefont {{Demorest}}\ \emph {et~al.}(2010)\citenamefont
  {{Demorest}}, \citenamefont {{Pennucci}}, \citenamefont {{Ransom}},
  \citenamefont {{Roberts}},\ and\ \citenamefont
  {{Hessels}}}]{2010Natur.467.1081D}%
  \BibitemOpen
  \bibfield  {author} {\bibinfo {author} {\bibfnamefont {P.~B.}\ \bibnamefont
  {{Demorest}}}, \bibinfo {author} {\bibfnamefont {T.}~\bibnamefont
  {{Pennucci}}}, \bibinfo {author} {\bibfnamefont {S.~M.}\ \bibnamefont
  {{Ransom}}}, \bibinfo {author} {\bibfnamefont {M.~S.~E.}\ \bibnamefont
  {{Roberts}}}, \ and\ \bibinfo {author} {\bibfnamefont {J.~W.~T.}\
  \bibnamefont {{Hessels}}},\ }\href {\doibase 10.1038/nature09466} {\bibfield
  {journal} {\bibinfo  {journal} {\nat}\ }\textbf {\bibinfo {volume} {467}},\
  \bibinfo {pages} {1081} (\bibinfo {year} {2010})},\ \Eprint
  {http://arxiv.org/abs/1010.5788} {arXiv:1010.5788 [astro-ph.HE]}\BibitemShut
  {NoStop}%
\bibitem [{\citenamefont {Fonseca}\ \emph {et~al.}(2016)\citenamefont {Fonseca}
  \emph {et~al.}}]{Fonseca:2016tux}%
  \BibitemOpen
  \bibfield  {author} {\bibinfo {author} {\bibfnamefont {E.}~\bibnamefont
  {Fonseca}} \emph {et~al.},\ }\href {\doibase 10.3847/0004-637X/832/2/167}
  {\bibfield  {journal} {\bibinfo  {journal} {Astrophys. J.}\ }\textbf
  {\bibinfo {volume} {832}},\ \bibinfo {pages} {167} (\bibinfo {year}
  {2016})},\ \Eprint {http://arxiv.org/abs/1603.00545} {arXiv:1603.00545
  [astro-ph.HE]}\BibitemShut {NoStop}%
\bibitem [{\citenamefont {{Antoniadis}}\ \emph {et~al.}(2013)\citenamefont
  {{Antoniadis}}, \citenamefont {{Freire}}, \citenamefont {{Wex}},
  \citenamefont {{Tauris}}, \citenamefont {{Lynch}}, \citenamefont {{van
  Kerkwijk}}, \citenamefont {{Kramer}}, \citenamefont {{Bassa}}, \citenamefont
  {{Dhillon}}, \citenamefont {{Driebe}}, \citenamefont {{Hessels}},
  \citenamefont {{Kaspi}}, \citenamefont {{Kondratiev}}, \citenamefont
  {{Langer}}, \citenamefont {{Marsh}}, \citenamefont {{McLaughlin}},
  \citenamefont {{Pennucci}}, \citenamefont {{Ransom}}, \citenamefont
  {{Stairs}}, \citenamefont {{van Leeuwen}}, \citenamefont {{Verbiest}},\ and\
  \citenamefont {{Whelan}}}]{2013Sci...340..448A}%
  \BibitemOpen
  \bibfield  {author} {\bibinfo {author} {\bibfnamefont {J.}~\bibnamefont
  {{Antoniadis}}}, \bibinfo {author} {\bibfnamefont {P.~C.~C.}\ \bibnamefont
  {{Freire}}}, \bibinfo {author} {\bibfnamefont {N.}~\bibnamefont {{Wex}}},
  \bibinfo {author} {\bibfnamefont {T.~M.}\ \bibnamefont {{Tauris}}}, \bibinfo
  {author} {\bibfnamefont {R.~S.}\ \bibnamefont {{Lynch}}}, \bibinfo {author}
  {\bibfnamefont {M.~H.}\ \bibnamefont {{van Kerkwijk}}}, \bibinfo {author}
  {\bibfnamefont {M.}~\bibnamefont {{Kramer}}}, \bibinfo {author}
  {\bibfnamefont {C.}~\bibnamefont {{Bassa}}}, \bibinfo {author} {\bibfnamefont
  {V.~S.}\ \bibnamefont {{Dhillon}}}, \bibinfo {author} {\bibfnamefont
  {T.}~\bibnamefont {{Driebe}}}, \bibinfo {author} {\bibfnamefont {J.~W.~T.}\
  \bibnamefont {{Hessels}}}, \bibinfo {author} {\bibfnamefont {V.~M.}\
  \bibnamefont {{Kaspi}}}, \bibinfo {author} {\bibfnamefont {V.~I.}\
  \bibnamefont {{Kondratiev}}}, \bibinfo {author} {\bibfnamefont
  {N.}~\bibnamefont {{Langer}}}, \bibinfo {author} {\bibfnamefont {T.~R.}\
  \bibnamefont {{Marsh}}}, \bibinfo {author} {\bibfnamefont {M.~A.}\
  \bibnamefont {{McLaughlin}}}, \bibinfo {author} {\bibfnamefont {T.~T.}\
  \bibnamefont {{Pennucci}}}, \bibinfo {author} {\bibfnamefont {S.~M.}\
  \bibnamefont {{Ransom}}}, \bibinfo {author} {\bibfnamefont {I.~H.}\
  \bibnamefont {{Stairs}}}, \bibinfo {author} {\bibfnamefont {J.}~\bibnamefont
  {{van Leeuwen}}}, \bibinfo {author} {\bibfnamefont {J.~P.~W.}\ \bibnamefont
  {{Verbiest}}}, \ and\ \bibinfo {author} {\bibfnamefont {D.~G.}\ \bibnamefont
  {{Whelan}}},\ }\href {\doibase 10.1126/science.1233232} {\bibfield  {journal}
  {\bibinfo  {journal} {Science}\ }\textbf {\bibinfo {volume} {340}},\ \bibinfo
  {pages} {448} (\bibinfo {year} {2013})},\ \Eprint
  {http://arxiv.org/abs/1304.6875} {arXiv:1304.6875 [astro-ph.HE]}\BibitemShut
  {NoStop}%
\bibitem [{\citenamefont {{Brambilla}}\ \emph {et~al.}(2014)\citenamefont
  {{Brambilla}}, \citenamefont {{Eidelman}}, \citenamefont {{Foka}},
  \citenamefont {{Gardner}}, \citenamefont {{Kronfeld}}, \citenamefont
  {{Alford}}, \citenamefont {{Alkofer}}, \citenamefont {{Butenschoen}},
  \citenamefont {{Cohen}}, \citenamefont {{Erdmenger}}, \citenamefont
  {{Fabbietti}}, \citenamefont {{Faber}}, \citenamefont {{Goity}},
  \citenamefont {{Ketzer}}, \citenamefont {{Lin}}, \citenamefont
  {{Llanes-Estrada}}, \citenamefont {{Meyer}}, \citenamefont {{Pakhlov}},
  \citenamefont {{Pallante}}, \citenamefont {{Polikarpov}}, \citenamefont
  {{Sazdjian}}, \citenamefont {{Schmitt}}, \citenamefont {{Snow}},
  \citenamefont {{Vairo}}, \citenamefont {{Vogt}}, \citenamefont {{Vuorinen}},
  \citenamefont {{Wittig}}, \citenamefont {{Arnold}}, \citenamefont
  {{Christakoglou}}, \citenamefont {{Di Nezza}}, \citenamefont {{Fodor}},
  \citenamefont {{Tormo}}, \citenamefont {{H{\"o}llwieser}}, \citenamefont
  {{Kalwait}}, \citenamefont {{Keane}}, \citenamefont {{Kiritsis}},
  \citenamefont {{Mischke}}, \citenamefont {{Mizuk}}, \citenamefont
  {{Odyniec}}, \citenamefont {{Papadodimas}}, \citenamefont {{Pich}},
  \citenamefont {{Pittau}}, \citenamefont {{Qiu}}, \citenamefont {{Ricciardi}},
  \citenamefont {{Salgado}}, \citenamefont {{Schwenzer}}, \citenamefont
  {{Stefanis}}, \citenamefont {{von Hippel}},\ and\ \citenamefont
  {{.~Zakharov}}}]{Brambilla:2014jmp}%
  \BibitemOpen
  \bibfield  {author} {\bibinfo {author} {\bibfnamefont {N.}~\bibnamefont
  {{Brambilla}}}, \bibinfo {author} {\bibfnamefont {S.}~\bibnamefont
  {{Eidelman}}}, \bibinfo {author} {\bibfnamefont {P.}~\bibnamefont {{Foka}}},
  \bibinfo {author} {\bibfnamefont {S.}~\bibnamefont {{Gardner}}}, \bibinfo
  {author} {\bibfnamefont {A.~S.}\ \bibnamefont {{Kronfeld}}}, \bibinfo
  {author} {\bibfnamefont {M.~G.}\ \bibnamefont {{Alford}}}, \bibinfo {author}
  {\bibfnamefont {R.}~\bibnamefont {{Alkofer}}}, \bibinfo {author}
  {\bibfnamefont {M.}~\bibnamefont {{Butenschoen}}}, \bibinfo {author}
  {\bibfnamefont {T.~D.}\ \bibnamefont {{Cohen}}}, \bibinfo {author}
  {\bibfnamefont {J.}~\bibnamefont {{Erdmenger}}}, \bibinfo {author}
  {\bibfnamefont {L.}~\bibnamefont {{Fabbietti}}}, \bibinfo {author}
  {\bibfnamefont {M.}~\bibnamefont {{Faber}}}, \bibinfo {author} {\bibfnamefont
  {J.~L.}\ \bibnamefont {{Goity}}}, \bibinfo {author} {\bibfnamefont
  {B.}~\bibnamefont {{Ketzer}}}, \bibinfo {author} {\bibfnamefont {H.~W.}\
  \bibnamefont {{Lin}}}, \bibinfo {author} {\bibfnamefont {F.~J.}\ \bibnamefont
  {{Llanes-Estrada}}}, \bibinfo {author} {\bibfnamefont {H.}~\bibnamefont
  {{Meyer}}}, \bibinfo {author} {\bibfnamefont {P.}~\bibnamefont {{Pakhlov}}},
  \bibinfo {author} {\bibfnamefont {E.}~\bibnamefont {{Pallante}}}, \bibinfo
  {author} {\bibfnamefont {M.~I.}\ \bibnamefont {{Polikarpov}}}, \bibinfo
  {author} {\bibfnamefont {H.}~\bibnamefont {{Sazdjian}}}, \bibinfo {author}
  {\bibfnamefont {A.}~\bibnamefont {{Schmitt}}}, \bibinfo {author}
  {\bibfnamefont {W.~M.}\ \bibnamefont {{Snow}}}, \bibinfo {author}
  {\bibfnamefont {A.}~\bibnamefont {{Vairo}}}, \bibinfo {author} {\bibfnamefont
  {R.}~\bibnamefont {{Vogt}}}, \bibinfo {author} {\bibfnamefont
  {A.}~\bibnamefont {{Vuorinen}}}, \bibinfo {author} {\bibfnamefont
  {H.}~\bibnamefont {{Wittig}}}, \bibinfo {author} {\bibfnamefont
  {P.}~\bibnamefont {{Arnold}}}, \bibinfo {author} {\bibfnamefont
  {P.}~\bibnamefont {{Christakoglou}}}, \bibinfo {author} {\bibfnamefont
  {P.}~\bibnamefont {{Di Nezza}}}, \bibinfo {author} {\bibfnamefont
  {Z.}~\bibnamefont {{Fodor}}}, \bibinfo {author} {\bibfnamefont {X.~G.~i.}\
  \bibnamefont {{Tormo}}}, \bibinfo {author} {\bibfnamefont {R.}~\bibnamefont
  {{H{\"o}llwieser}}}, \bibinfo {author} {\bibfnamefont {A.}~\bibnamefont
  {{Kalwait}}}, \bibinfo {author} {\bibfnamefont {D.}~\bibnamefont {{Keane}}},
  \bibinfo {author} {\bibfnamefont {E.}~\bibnamefont {{Kiritsis}}}, \bibinfo
  {author} {\bibfnamefont {A.}~\bibnamefont {{Mischke}}}, \bibinfo {author}
  {\bibfnamefont {R.}~\bibnamefont {{Mizuk}}}, \bibinfo {author} {\bibfnamefont
  {G.}~\bibnamefont {{Odyniec}}}, \bibinfo {author} {\bibfnamefont
  {K.}~\bibnamefont {{Papadodimas}}}, \bibinfo {author} {\bibfnamefont
  {A.}~\bibnamefont {{Pich}}}, \bibinfo {author} {\bibfnamefont
  {R.}~\bibnamefont {{Pittau}}}, \bibinfo {author} {\bibfnamefont {J.-W.}\
  \bibnamefont {{Qiu}}}, \bibinfo {author} {\bibfnamefont {G.}~\bibnamefont
  {{Ricciardi}}}, \bibinfo {author} {\bibfnamefont {C.~A.}\ \bibnamefont
  {{Salgado}}}, \bibinfo {author} {\bibfnamefont {K.}~\bibnamefont
  {{Schwenzer}}}, \bibinfo {author} {\bibfnamefont {N.~G.}\ \bibnamefont
  {{Stefanis}}}, \bibinfo {author} {\bibfnamefont {G.~M.}\ \bibnamefont {{von
  Hippel}}}, \ and\ \bibinfo {author} {\bibfnamefont {V.~I.}\ \bibnamefont
  {{.~Zakharov}}},\ }\href@noop {} {\bibfield  {journal} {\bibinfo  {journal}
  {Eur.\ Phys.\ J.\ C}\ }\textbf {\bibinfo {volume} {74}},\ \bibinfo {pages}
  {2981} (\bibinfo {year} {2014})},\ \Eprint {http://arxiv.org/abs/1404.3723}
  {1404.3723 [hep-ph]}\BibitemShut {NoStop}%
\bibitem [{\citenamefont {{Buballa}}\ \emph {et~al.}(2014)\citenamefont
  {{Buballa}}, \citenamefont {{Dexheimer}}, \citenamefont {{Drago}},
  \citenamefont {{Fraga}}, \citenamefont {{Haensel}}, \citenamefont
  {{Mishustin}}, \citenamefont {{Pagliara}}, \citenamefont
  {{Schaffner-Bielich}}, \citenamefont {{Schramm}}, \citenamefont
  {{Sedrakian}},\ and\ \citenamefont {{Weber}}}]{2014JPhG...41l3001B}%
  \BibitemOpen
  \bibfield  {author} {\bibinfo {author} {\bibfnamefont {M.}~\bibnamefont
  {{Buballa}}}, \bibinfo {author} {\bibfnamefont {V.}~\bibnamefont
  {{Dexheimer}}}, \bibinfo {author} {\bibfnamefont {A.}~\bibnamefont
  {{Drago}}}, \bibinfo {author} {\bibfnamefont {E.}~\bibnamefont {{Fraga}}},
  \bibinfo {author} {\bibfnamefont {P.}~\bibnamefont {{Haensel}}}, \bibinfo
  {author} {\bibfnamefont {I.}~\bibnamefont {{Mishustin}}}, \bibinfo {author}
  {\bibfnamefont {G.}~\bibnamefont {{Pagliara}}}, \bibinfo {author}
  {\bibfnamefont {J.}~\bibnamefont {{Schaffner-Bielich}}}, \bibinfo {author}
  {\bibfnamefont {S.}~\bibnamefont {{Schramm}}}, \bibinfo {author}
  {\bibfnamefont {A.}~\bibnamefont {{Sedrakian}}}, \ and\ \bibinfo {author}
  {\bibfnamefont {F.}~\bibnamefont {{Weber}}},\ }\href {\doibase
  10.1088/0954-3899/41/12/123001} {\bibfield  {journal} {\bibinfo  {journal}
  {J.  Phys. G }\ }\textbf {\bibinfo {volume} {41}},\
  \bibinfo {eid} {123001} (\bibinfo {year} {2014})},\ \Eprint
  {http://arxiv.org/abs/1402.6911} {arXiv:1402.6911 [astro-ph.HE]}\BibitemShut
  {NoStop}%
\bibitem [{\citenamefont {{Alford}}\ \emph {et~al.}(2008)\citenamefont
  {{Alford}}, \citenamefont {{Schmitt}}, \citenamefont {{Rajagopal}},\ and\
  \citenamefont {{Sch{\"a}fer}}}]{2008RvMP...80.1455A}%
  \BibitemOpen
  \bibfield  {author} {\bibinfo {author} {\bibfnamefont {M.~G.}\ \bibnamefont
  {{Alford}}}, \bibinfo {author} {\bibfnamefont {A.}~\bibnamefont {{Schmitt}}},
  \bibinfo {author} {\bibfnamefont {K.}~\bibnamefont {{Rajagopal}}}, \ and\
  \bibinfo {author} {\bibfnamefont {T.}~\bibnamefont {{Sch{\"a}fer}}},\ }\href
  {\doibase 10.1103/RevModPhys.80.1455} {\bibfield  {journal} {\bibinfo
  {journal} {Rev.  Mod. Phys.}\ }\textbf {\bibinfo {volume} {80}},\
  \bibinfo {pages} {1455} (\bibinfo {year} {2008})},\ \Eprint
  {http://arxiv.org/abs/0709.4635} {arXiv:0709.4635 [hep-ph]}\BibitemShut
  {NoStop}%
\bibitem [{\citenamefont {{Anglani}}\ \emph {et~al.}(2014)\citenamefont
  {{Anglani}}, \citenamefont {{Casalbuoni}}, \citenamefont {{Ciminale}},
  \citenamefont {{Ippolito}}, \citenamefont {{Gatto}}, \citenamefont
  {{Mannarelli}},\ and\ \citenamefont {{Ruggieri}}}]{2014RvMP...86..509A}%
  \BibitemOpen
  \bibfield  {author} {\bibinfo {author} {\bibfnamefont {R.}~\bibnamefont
  {{Anglani}}}, \bibinfo {author} {\bibfnamefont {R.}~\bibnamefont
  {{Casalbuoni}}}, \bibinfo {author} {\bibfnamefont {M.}~\bibnamefont
  {{Ciminale}}}, \bibinfo {author} {\bibfnamefont {N.}~\bibnamefont
  {{Ippolito}}}, \bibinfo {author} {\bibfnamefont {R.}~\bibnamefont {{Gatto}}},
  \bibinfo {author} {\bibfnamefont {M.}~\bibnamefont {{Mannarelli}}}, \ and\
  \bibinfo {author} {\bibfnamefont {M.}~\bibnamefont {{Ruggieri}}},\ }\href
  {\doibase 10.1103/RevModPhys.86.509} {\bibfield  {journal} {\bibinfo
  {journal} {Rev. Mod. Phys.}\ }\textbf {\bibinfo {volume} {86}},\
  \bibinfo {pages} {509} (\bibinfo {year} {2014})},\ \Eprint
  {http://arxiv.org/abs/1302.4264} {arXiv:1302.4264 [hep-ph]}\BibitemShut
  {NoStop}%
\bibitem [{\citenamefont {Ruster}\ \emph {et~al.}(2005)\citenamefont
  {R\"uster}, \citenamefont {Werth}, \citenamefont {Buballa}, \citenamefont
  {Shovkovy},\ and\ \citenamefont {Rischke}}]{Ruester:2005jc}%
  \BibitemOpen
  \bibfield  {author} {\bibinfo {author} {\bibfnamefont {S.~B.}\ \bibnamefont
  {R\"uster}}, \bibinfo {author} {\bibfnamefont {V.}~\bibnamefont {Werth}},
  \bibinfo {author} {\bibfnamefont {M.}~\bibnamefont {Buballa}}, \bibinfo
  {author} {\bibfnamefont {I.~A.}\ \bibnamefont {Shovkovy}}, \ and\ \bibinfo
  {author} {\bibfnamefont {D.~H.}\ \bibnamefont {Rischke}},\ }\href {\doibase
  10.1103/PhysRevD.72.034004} {\bibfield  {journal} {\bibinfo  {journal} {Phys.
  Rev.}\ }\textbf {\bibinfo {volume} {D72}},\ \bibinfo {pages} {034004}
  (\bibinfo {year} {2005})},\ \Eprint {http://arxiv.org/abs/hep-ph/0503184}
  {arXiv:hep-ph/0503184 [hep-ph]}\BibitemShut {NoStop}%
\bibitem [{\citenamefont {{Blaschke}}\ \emph {et~al.}(2005)\citenamefont
  {{Blaschke}}, \citenamefont {{Fredriksson}}, \citenamefont {{Grigorian}},
  \citenamefont {{{\"O}zta{\c s}}},\ and\ \citenamefont
  {{Sandin}}}]{2005PhRvD..72f5020B}%
  \BibitemOpen
  \bibfield  {author} {\bibinfo {author} {\bibfnamefont {D.}~\bibnamefont
  {{Blaschke}}}, \bibinfo {author} {\bibfnamefont {S.}~\bibnamefont
  {{Fredriksson}}}, \bibinfo {author} {\bibfnamefont {H.}~\bibnamefont
  {{Grigorian}}}, \bibinfo {author} {\bibfnamefont {A.~M.}\ \bibnamefont
  {{{\"O}zta{\c s}}}}, \ and\ \bibinfo {author} {\bibfnamefont
  {F.}~\bibnamefont {{Sandin}}},\ }\href {\doibase 10.1103/PhysRevD.72.065020}
  {\bibfield  {journal} {\bibinfo  {journal} {\prd}\ }\textbf {\bibinfo
  {volume} {72}},\ \bibinfo {eid} {065020} (\bibinfo {year} {2005})},\ \Eprint
  {http://arxiv.org/abs/hep-ph/0503194} {hep-ph/0503194}\BibitemShut {NoStop}%
\bibitem [{\citenamefont {{Blaschke}}\ \emph {et~al.}(2009)\citenamefont
  {{Blaschke}}, \citenamefont {{Sandin}}, \citenamefont {{Kl{\"a}hn}},\ and\
  \citenamefont {{Berdermann}}}]{2009PhRvC..80f5807B}%
  \BibitemOpen
  \bibfield  {author} {\bibinfo {author} {\bibfnamefont {D.}~\bibnamefont
  {{Blaschke}}}, \bibinfo {author} {\bibfnamefont {F.}~\bibnamefont
  {{Sandin}}}, \bibinfo {author} {\bibfnamefont {T.}~\bibnamefont
  {{Kl{\"a}hn}}}, \ and\ \bibinfo {author} {\bibfnamefont {J.}~\bibnamefont
  {{Berdermann}}},\ }\href {\doibase 10.1103/PhysRevC.80.065807} {\bibfield
  {journal} {\bibinfo  {journal} {\prc}\ }\textbf {\bibinfo {volume} {80}},\
  \bibinfo {eid} {065807} (\bibinfo {year} {2009})},\ \Eprint
  {http://arxiv.org/abs/0807.0414} {arXiv:0807.0414 [nucl-th]}\BibitemShut
  {NoStop}%
\bibitem [{\citenamefont {{Bonanno}}\ and\ \citenamefont
  {{Sedrakian}}(2012)}]{2012A&A...539A..16B}%
  \BibitemOpen
  \bibfield  {author} {\bibinfo {author} {\bibfnamefont {L.}~\bibnamefont
  {{Bonanno}}}\ and\ \bibinfo {author} {\bibfnamefont {A.}~\bibnamefont
  {{Sedrakian}}},\ }\href {\doibase 10.1051/0004-6361/201117832} {\bibfield
  {journal} {\bibinfo  {journal} {\aap}\ }\textbf {\bibinfo {volume} {539}},\
  \bibinfo {eid} {A16} (\bibinfo {year} {2012})},\ \Eprint
  {http://arxiv.org/abs/1108.0559} {arXiv:1108.0559 [astro-ph.SR]}\BibitemShut
  {NoStop}%
\bibitem [{\citenamefont {Warringa}(2006)}]{Warringa:2006dk}%
  \BibitemOpen
  \bibfield  {author} {\bibinfo {author} {\bibfnamefont {H.~J.}\ \bibnamefont
  {Warringa}},\ }\href@noop {} {\  (\bibinfo {year} {2006})},\ \Eprint
  {http://arxiv.org/abs/hep-ph/0606063} {arXiv:hep-ph/0606063 [hep-ph]}\BibitemShut {NoStop}%
\bibitem [{\citenamefont {Fukushima}\ and\ \citenamefont
  {Hatsuda}(2011)}]{Fukushima:2010bq}%
  \BibitemOpen
  \bibfield  {author} {\bibinfo {author} {\bibfnamefont {K.}~\bibnamefont
  {Fukushima}}\ and\ \bibinfo {author} {\bibfnamefont {T.}~\bibnamefont
  {Hatsuda}},\ }\href {\doibase 10.1088/0034-4885/74/1/014001} {\bibfield
  {journal} {\bibinfo  {journal} {Rep. Prog. Phys.}\ }\textbf {\bibinfo
  {volume} {74}},\ \bibinfo {pages} {014001} (\bibinfo {year} {2011})},\
  \Eprint {http://arxiv.org/abs/1005.4814} {arXiv:1005.4814 [hep-ph]}
 \BibitemShut {NoStop}%
\bibitem [{\citenamefont {{Buballa}}\ \emph {et~al.}(2004)\citenamefont
  {{Buballa}}, \citenamefont {{Neumann}}, \citenamefont {{Oertel}},\ and\
  \citenamefont {{Shovkovy}}}]{2004PhLB..595...36B}%
  \BibitemOpen
  \bibfield  {author} {\bibinfo {author} {\bibfnamefont {M.}~\bibnamefont
  {{Buballa}}}, \bibinfo {author} {\bibfnamefont {F.}~\bibnamefont
  {{Neumann}}}, \bibinfo {author} {\bibfnamefont {M.}~\bibnamefont {{Oertel}}},
  \ and\ \bibinfo {author} {\bibfnamefont {I.}~\bibnamefont {{Shovkovy}}},\
  }\href {\doibase 10.1016/j.physletb.2004.05.064} {\bibfield  {journal}
  {\bibinfo  {journal} {Phys. Lett. B}\ }\textbf {\bibinfo {volume}
  {595}},\ \bibinfo {pages} {36} (\bibinfo {year} {2004})},\ \Eprint
  {http://arxiv.org/abs/nucl-th/0312078} {nucl-th/0312078}\BibitemShut
  {NoStop}%
\bibitem [{\citenamefont {{Kl{\"a}hn}}\ \emph {et~al.}(2007)\citenamefont
  {{Kl{\"a}hn}}, \citenamefont {{Blaschke}}, \citenamefont {{Sandin}},
  \citenamefont {{Fuchs}}, \citenamefont {{Faessler}}, \citenamefont
  {{Grigorian}}, \citenamefont {{R{\"o}pke}},\ and\ \citenamefont
  {{Tr{\"u}mper}}}]{2007PhLB..654..170K}%
  \BibitemOpen
  \bibfield  {author} {\bibinfo {author} {\bibfnamefont {T.}~\bibnamefont
  {{Kl{\"a}hn}}}, \bibinfo {author} {\bibfnamefont {D.}~\bibnamefont
  {{Blaschke}}}, \bibinfo {author} {\bibfnamefont {F.}~\bibnamefont
  {{Sandin}}}, \bibinfo {author} {\bibfnamefont {C.}~\bibnamefont {{Fuchs}}},
  \bibinfo {author} {\bibfnamefont {A.}~\bibnamefont {{Faessler}}}, \bibinfo
  {author} {\bibfnamefont {H.}~\bibnamefont {{Grigorian}}}, \bibinfo {author}
  {\bibfnamefont {G.}~\bibnamefont {{R{\"o}pke}}}, \ and\ \bibinfo {author}
  {\bibfnamefont {J.}~\bibnamefont {{Tr{\"u}mper}}},\ }\href {\doibase
  10.1016/j.physletb.2007.08.048} {\bibfield  {journal} {\bibinfo  {journal}
  {Phys. Lett. B}\ }\textbf {\bibinfo {volume} {654}},\ \bibinfo {pages}
  {170} (\bibinfo {year} {2007})},\ \Eprint
  {http://arxiv.org/abs/nucl-th/0609067} {nucl-th/0609067}\BibitemShut
  {NoStop}%
\bibitem [{\citenamefont {{Kl{\"a}hn}}\ \emph {et~al.}(2013)\citenamefont
  {{Kl{\"a}hn}}, \citenamefont {{{\L}astowiecki}},\ and\ \citenamefont
  {{Blaschke}}}]{2013PhRvD..88h5001K}%
  \BibitemOpen
  \bibfield  {author} {\bibinfo {author} {\bibfnamefont {T.}~\bibnamefont
  {{Kl{\"a}hn}}}, \bibinfo {author} {\bibfnamefont {R.}~\bibnamefont
  {{{\L}astowiecki}}}, \ and\ \bibinfo {author} {\bibfnamefont
  {D.}~\bibnamefont {{Blaschke}}},\ }\href {\doibase
  10.1103/PhysRevD.88.085001} {\bibfield  {journal} {\bibinfo  {journal}
  {\prd}\ }\textbf {\bibinfo {volume} {88}},\ \bibinfo {eid} {085001} (\bibinfo
  {year} {2013})},\ \Eprint {http://arxiv.org/abs/1307.6996} {arXiv:1307.6996
  [nucl-th]}\BibitemShut {NoStop}%
\bibitem [{\citenamefont {{Dexheimer}}\ \emph {et~al.}(2015)\citenamefont
  {{Dexheimer}}, \citenamefont {{Negreiros}},\ and\ \citenamefont
  {{Schramm}}}]{2015PhRvC..91e5808D}%
  \BibitemOpen
  \bibfield  {author} {\bibinfo {author} {\bibfnamefont {V.}~\bibnamefont
  {{Dexheimer}}}, \bibinfo {author} {\bibfnamefont {R.}~\bibnamefont
  {{Negreiros}}}, \ and\ \bibinfo {author} {\bibfnamefont {S.}~\bibnamefont
  {{Schramm}}},\ }\href {\doibase 10.1103/PhysRevC.91.055808} {\bibfield
  {journal} {\bibinfo  {journal} {\prc}\ }\textbf {\bibinfo {volume} {91}},\
  \bibinfo {eid} {055808} (\bibinfo {year} {2015})},\ \Eprint
  {http://arxiv.org/abs/1411.4623} {arXiv:1411.4623 [astro-ph.HE]}\BibitemShut
  {NoStop}%
\bibitem [{\citenamefont {{Alvarez-Castillo}}\ \emph
  {et~al.}(2016{\natexlab{a}})\citenamefont {{Alvarez-Castillo}}, \citenamefont
  {{Beni{\'c}}}, \citenamefont {{Blaschke}}, \citenamefont {{Han}},\ and\
  \citenamefont {{Typel}}}]{2016EPJA...52..232A}%
  \BibitemOpen
  \bibfield  {author} {\bibinfo {author} {\bibfnamefont {D.}~\bibnamefont
  {{Alvarez-Castillo}}}, \bibinfo {author} {\bibfnamefont {S.}~\bibnamefont
  {{Beni{\'c}}}}, \bibinfo {author} {\bibfnamefont {D.}~\bibnamefont
  {{Blaschke}}}, \bibinfo {author} {\bibfnamefont {S.}~\bibnamefont {{Han}}}, \
  and\ \bibinfo {author} {\bibfnamefont {S.}~\bibnamefont {{Typel}}},\ }\href
  {\doibase 10.1140/epja/i2016-16232-9} {\bibfield  {journal} {\bibinfo
  {journal} {Eur. Phys. J. A}\ }\textbf {\bibinfo {volume} {52}},\
  \bibinfo {eid} {232} (\bibinfo {year} {2016}{\natexlab{a}})},\ \Eprint
  {http://arxiv.org/abs/1608.02425} {arXiv:1608.02425 [nucl-th]}\BibitemShut
  {NoStop}%
\bibitem [{\citenamefont {Alvarez-Castillo}\ and\ \citenamefont
  {Blaschke}(2015)}]{Alvarez-Castillo:2014dva}%
  \BibitemOpen
  \bibfield  {author} {\bibinfo {author} {\bibfnamefont {D.~E.}\ \bibnamefont
  {Alvarez-Castillo}}\ and\ \bibinfo {author} {\bibfnamefont {D.}~\bibnamefont
  {Blaschke}},\ }{\bibfield  {journal} {\bibinfo  {journal} {Phys.
  Part. Nucl.}\ }\textbf {\bibinfo {volume} {46}},\ \bibinfo {pages} {846}
  (\bibinfo {year} {2015})},\ \Eprint {http://arxiv.org/abs/1412.8463}
  {arXiv:1412.8463 [astro-ph.HE]}\BibitemShut {NoStop}%
\bibitem [{\citenamefont {{Alvarez-Castillo}}\ \emph
  {et~al.}(2016{\natexlab{b}})\citenamefont {{Alvarez-Castillo}}, \citenamefont
  {{Ayriyan}}, \citenamefont {{Beni{\'c}}}, \citenamefont {{Blaschke}},
  \citenamefont {{Grigorian}},\ and\ \citenamefont
  {{Typel}}}]{2016EPJA...52...69A}%
  \BibitemOpen
  \bibfield  {author} {\bibinfo {author} {\bibfnamefont {D.}~\bibnamefont
  {{Alvarez-Castillo}}}, \bibinfo {author} {\bibfnamefont {A.}~\bibnamefont
  {{Ayriyan}}}, \bibinfo {author} {\bibfnamefont {S.}~\bibnamefont
  {{Beni{\'c}}}}, \bibinfo {author} {\bibfnamefont {D.}~\bibnamefont
  {{Blaschke}}}, \bibinfo {author} {\bibfnamefont {H.}~\bibnamefont
  {{Grigorian}}}, \ and\ \bibinfo {author} {\bibfnamefont {S.}~\bibnamefont
  {{Typel}}},\ }\href {\doibase 10.1140/epja/i2016-16069-2} {\bibfield
  {journal} {\bibinfo  {journal} {Eur. Phys. J. A}\ }\textbf
  {\bibinfo {volume} {52}},\ \bibinfo {eid} {69} (\bibinfo {year}
  {2016}{\natexlab{b}})},\ \Eprint {http://arxiv.org/abs/1603.03457}
  {arXiv:1603.03457 [nucl-th]}\BibitemShut {NoStop}%
\bibitem [{\citenamefont {{Zacchi}}\ \emph {et~al.}(2017)\citenamefont
  {{Zacchi}}, \citenamefont {{Tolos}},\ and\ \citenamefont
  {{Schaffner-Bielich}}}]{2017PhRvD..95j3008Z}%
  \BibitemOpen
  \bibfield  {author} {\bibinfo {author} {\bibfnamefont {A.}~\bibnamefont
  {{Zacchi}}}, \bibinfo {author} {\bibfnamefont {L.}~\bibnamefont {{Tolos}}}, \
  and\ \bibinfo {author} {\bibfnamefont {J.}~\bibnamefont
  {{Schaffner-Bielich}}},\ }\href {\doibase 10.1103/PhysRevD.95.103008}
  {\bibfield  {journal} {\bibinfo  {journal} {\prd}\ }\textbf {\bibinfo
  {volume} {95}},\ \bibinfo {eid} {103008} (\bibinfo {year} {2017})},\ \Eprint
  {http://arxiv.org/abs/1612.06167} {arXiv:1612.06167 [astro-ph.HE]}
 \BibitemShut {NoStop}%
\bibitem [{\citenamefont {{Agrawal}}\ and\ \citenamefont
  {{Dhiman}}(2009)}]{2009PhRvD..79j3006A}%
  \BibitemOpen
  \bibfield  {author} {\bibinfo {author} {\bibfnamefont {B.~K.}\ \bibnamefont
  {{Agrawal}}}\ and\ \bibinfo {author} {\bibfnamefont {S.~K.}\ \bibnamefont
  {{Dhiman}}},\ }\href {\doibase 10.1103/PhysRevD.79.103006} {\bibfield
  {journal} {\bibinfo  {journal} {\prd}\ }\textbf {\bibinfo {volume} {79}},\
  \bibinfo {eid} {103006} (\bibinfo {year} {2009})},\ \Eprint
  {http://arxiv.org/abs/0904.2946} {arXiv:0904.2946 [astro-ph.HE]}\BibitemShut
  {NoStop}%
\bibitem [{\citenamefont {{Beni{\'c}}}\ \emph {et~al.}(2015)\citenamefont
  {{Beni{\'c}}}, \citenamefont {{Blaschke}}, \citenamefont
  {{Alvarez-Castillo}}, \citenamefont {{Fischer}},\ and\ \citenamefont
  {{Typel}}}]{2015A&A...577A..40B}%
  \BibitemOpen
  \bibfield  {author} {\bibinfo {author} {\bibfnamefont {S.}~\bibnamefont
  {{Beni{\'c}}}}, \bibinfo {author} {\bibfnamefont {D.}~\bibnamefont
  {{Blaschke}}}, \bibinfo {author} {\bibfnamefont {D.~E.}\ \bibnamefont
  {{Alvarez-Castillo}}}, \bibinfo {author} {\bibfnamefont {T.}~\bibnamefont
  {{Fischer}}}, \ and\ \bibinfo {author} {\bibfnamefont {S.}~\bibnamefont
  {{Typel}}},\ }\href {\doibase 10.1051/0004-6361/201425318} {\bibfield
  {journal} {\bibinfo  {journal} {\aap}\ }\textbf {\bibinfo {volume} {577}},\
  \bibinfo {eid} {A40} (\bibinfo {year} {2015})},\ \Eprint
  {http://arxiv.org/abs/1411.2856} {arXiv:1411.2856 [astro-ph.HE]}\BibitemShut
  {NoStop}%
\bibitem [{\citenamefont {{Blaschke}}\ \emph {et~al.}(2013)\citenamefont
  {{Blaschke}}, \citenamefont {{Alvarez-Castillo}},\ and\ \citenamefont
  {{Beni{\'c}}}}]{BlaPOS}%
  \BibitemOpen
  \bibfield  {author} {\bibinfo {author} {\bibfnamefont {D.}~\bibnamefont
  {{Blaschke}}}, \bibinfo {author} {\bibfnamefont {D.~E.}\ \bibnamefont
  {{Alvarez-Castillo}}}, \ and\ \bibinfo {author} {\bibfnamefont
  {S.}~\bibnamefont {{Beni{\'c}}}},\ }\href@noop {} {\bibfield  {journal}
  {\bibinfo  {journal} {Proc.Sci. CPOD}\ }\textbf {\bibinfo {volume} {063}} (\bibinfo
  {year} {2013})}\BibitemShut {NoStop}%
\bibitem [{\citenamefont {Alford}\ \emph {et~al.}(2001)\citenamefont {Alford},
  \citenamefont {Rajagopal}, \citenamefont {Reddy},\ and\ \citenamefont
  {Wilczek}}]{Alford:2001zr}%
  \BibitemOpen
  \bibfield  {author} {\bibinfo {author} {\bibfnamefont {M.~G.}\ \bibnamefont
  {Alford}}, \bibinfo {author} {\bibfnamefont {K.}~\bibnamefont {Rajagopal}},
  \bibinfo {author} {\bibfnamefont {S.}~\bibnamefont {Reddy}}, \ and\ \bibinfo
  {author} {\bibfnamefont {F.}~\bibnamefont {Wilczek}},\ }\href {\doibase
  10.1103/PhysRevD.64.074017} {\bibfield  {journal} {\bibinfo  {journal}
  {Phys.Rev.}\ }\textbf {\bibinfo {volume} {D64}},\ \bibinfo {pages} {074017}
  (\bibinfo {year} {2001})},\ \Eprint {http://arxiv.org/abs/hep-ph/0105009}
  {arXiv:hep-ph/0105009 [hep-ph]}\BibitemShut {NoStop}%
\bibitem [{\citenamefont {Palhares}\ and\ \citenamefont
  {Fraga}(2010)}]{Palhares:2010be}%
  \BibitemOpen
  \bibfield  {author} {\bibinfo {author} {\bibfnamefont {L.~F.}\ \bibnamefont
  {Palhares}}\ and\ \bibinfo {author} {\bibfnamefont {E.~S.}\ \bibnamefont
  {Fraga}},\ }\href {\doibase 10.1103/PhysRevD.82.125018} {\bibfield  {journal}
  {\bibinfo  {journal} {Phys.Rev.}\ }\textbf {\bibinfo {volume} {D82}},\
  \bibinfo {pages} {125018} (\bibinfo {year} {2010})},\ \Eprint
  {http://arxiv.org/abs/1006.2357} {arXiv:1006.2357 [hep-ph]}\BibitemShut
  {NoStop}%
\bibitem [{\citenamefont {Pinto}\ \emph {et~al.}(2012)\citenamefont {Pinto},
  \citenamefont {Koch},\ and\ \citenamefont {Randrup}}]{Pinto:2012aq}%
  \BibitemOpen
  \bibfield  {author} {\bibinfo {author} {\bibfnamefont {M.~B.}\ \bibnamefont
  {Pinto}}, \bibinfo {author} {\bibfnamefont {V.}~\bibnamefont {Koch}}, \ and\
  \bibinfo {author} {\bibfnamefont {J.}~\bibnamefont {Randrup}},\ }\href
  {\doibase 10.1103/PhysRevC.86.025203} {\bibfield  {journal} {\bibinfo
  {journal} {Phys. Rev.}\ }\textbf {\bibinfo {volume} {C86}},\ \bibinfo {pages}
  {025203} (\bibinfo {year} {2012})},\ \Eprint {http://arxiv.org/abs/1207.5186}
  {arXiv:1207.5186 [hep-ph]}\BibitemShut {NoStop}%
\bibitem [{\citenamefont {{Alford}}\ \emph {et~al.}(2013)\citenamefont
  {{Alford}}, \citenamefont {{Han}},\ and\ \citenamefont
  {{Prakash}}}]{2013PhRvD..88h3013A}%
  \BibitemOpen
  \bibfield  {author} {\bibinfo {author} {\bibfnamefont {M.~G.}\ \bibnamefont
  {{Alford}}}, \bibinfo {author} {\bibfnamefont {S.}~\bibnamefont {{Han}}}, \
  and\ \bibinfo {author} {\bibfnamefont {M.}~\bibnamefont {{Prakash}}},\ }\href
  {\doibase 10.1103/PhysRevD.88.083013} {\bibfield  {journal} {\bibinfo
  {journal} {\prd}\ }\textbf {\bibinfo {volume} {88}},\ \bibinfo {eid} {083013}
  (\bibinfo {year} {2013})},\ \Eprint {http://arxiv.org/abs/1302.4732}
  {arXiv:1302.4732 [astro-ph.SR]}\BibitemShut {NoStop}%
\bibitem [{\citenamefont {{Seidov}}(1971)}]{1971SvA....15..347S}%
  \BibitemOpen
  \bibfield  {author} {\bibinfo {author} {\bibfnamefont {Z.~F.}\ \bibnamefont
  {{Seidov}}},\ }\href@noop {} {\bibfield  {journal} {\bibinfo  {journal}
  {\sovast}\ }\textbf {\bibinfo {volume} {15}},\ \bibinfo {pages} {347}
  (\bibinfo {year} {1971})}\BibitemShut {NoStop}%
\bibitem [{\citenamefont {{Zdunik}}\ and\ \citenamefont
  {{Haensel}}(2013)}]{2013A&A...551A..61Z}%
  \BibitemOpen
  \bibfield  {author} {\bibinfo {author} {\bibfnamefont {J.~L.}\ \bibnamefont
  {{Zdunik}}}\ and\ \bibinfo {author} {\bibfnamefont {P.}~\bibnamefont
  {{Haensel}}},\ }\href {\doibase 10.1051/0004-6361/201220697} {\bibfield
  {journal} {\bibinfo  {journal} {\aap}\ }\textbf {\bibinfo {volume} {551}},\
  \bibinfo {eid} {A61} (\bibinfo {year} {2013})},\ \Eprint
  {http://arxiv.org/abs/1211.1231} {arXiv:1211.1231 [astro-ph.SR]}\BibitemShut
  {NoStop}%
\bibitem [{\citenamefont {{Colucci}}\ and\ \citenamefont
  {{Sedrakian}}(2013)}]{2013PhRvC..87e5806C}%
  \BibitemOpen
  \bibfield  {author} {\bibinfo {author} {\bibfnamefont {G.}~\bibnamefont
  {{Colucci}}}\ and\ \bibinfo {author} {\bibfnamefont {A.}~\bibnamefont
  {{Sedrakian}}},\ }\href {\doibase 10.1103/PhysRevC.87.055806} {\bibfield
  {journal} {\bibinfo  {journal} {\prc}\ }\textbf {\bibinfo {volume} {87}},\
  \bibinfo {eid} {055806} (\bibinfo {year} {2013})},\ \Eprint
  {http://arxiv.org/abs/1302.6925} {arXiv:1302.6925 [nucl-th]}\BibitemShut
  {NoStop}%
\bibitem{Fortin:2016hny} 
  M.~Fortin, C.~Providencia, A.~R.~Raduta, F.~Gulminelli, J.~L.~Zdunik, P.~Haensel and M.~Bejger,
  \textcolor{blue}{Phys.\ Rev.\ C {\bf 94}, 035804 (2016)
[arXiv:1604.01944 [astro-ph.SR]]}.
\bibitem [{\citenamefont {Tolman}(1939)}]{Tolman}%
  \BibitemOpen
  \bibfield  {author} {\bibinfo {author} {\bibfnamefont {R.~C.}\ \bibnamefont
  {Tolman}},\ }\href {\doibase 10.1103/PhysRev.55.364} {\bibfield  {journal}
  {\bibinfo  {journal} {Phys. Rev.}\ }\textbf {\bibinfo {volume} {55}},\
  \bibinfo {pages} {364} (\bibinfo {year} {1939})}\BibitemShut {NoStop}%
\bibitem [{\citenamefont {Oppenheimer}\ and\ \citenamefont
  {Volkoff}(1939)}]{OV}%
  \BibitemOpen
  \bibfield  {author} {\bibinfo {author} {\bibfnamefont {J.~R.}\ \bibnamefont
  {Oppenheimer}}\ and\ \bibinfo {author} {\bibfnamefont {G.~M.}\ \bibnamefont
  {Volkoff}},\ }\href {\doibase 10.1103/PhysRev.55.374} {\bibfield  {journal}
  {\bibinfo  {journal} {Phys. Rev.}\ }\textbf {\bibinfo {volume} {55}},\
  \bibinfo {pages} {374} (\bibinfo {year} {1939})}\BibitemShut {NoStop}%
\bibitem [{\citenamefont {{Bardeen}}\ \emph {et~al.}(1966)\citenamefont
  {{Bardeen}}, \citenamefont {{Thorne}},\ and\ \citenamefont
  {{Meltzer}}}]{BTM_methods}%
  \BibitemOpen
  \bibfield  {author} {\bibinfo {author} {\bibfnamefont {J.~M.}\ \bibnamefont
  {{Bardeen}}}, \bibinfo {author} {\bibfnamefont {K.~S.}\ \bibnamefont
  {{Thorne}}}, \ and\ \bibinfo {author} {\bibfnamefont {D.~W.}\ \bibnamefont
  {{Meltzer}}},\ }\href {\doibase 10.1086/148791} {\bibfield  {journal}
  {\bibinfo  {journal} {\apj}\ }\textbf {\bibinfo {volume} {145}},\ \bibinfo
  {pages} {505} (\bibinfo {year} {1966})}\BibitemShut {NoStop}%
\bibitem [{\citenamefont {{Glendenning}}\ \emph {et~al.}(1997)\citenamefont
  {{Glendenning}}, \citenamefont {{Pei}},\ and\ \citenamefont
  {{Weber}}}]{1997PhRvL..79.1603G}%
  \BibitemOpen
  \bibfield  {author} {\bibinfo {author} {\bibfnamefont {N.~K.}\ \bibnamefont
  {{Glendenning}}}, \bibinfo {author} {\bibfnamefont {S.}~\bibnamefont
  {{Pei}}}, \ and\ \bibinfo {author} {\bibfnamefont {F.}~\bibnamefont
  {{Weber}}},\ }\href {\doibase 10.1103/PhysRevLett.79.1603} {\bibfield
  {journal} {\bibinfo  {journal} {Phys. Rev. Lett.}\ }\textbf {\bibinfo
  {volume} {79}},\ \bibinfo {pages} {1603} (\bibinfo {year} {1997})},\ \Eprint
  {http://arxiv.org/abs/astro-ph/9705235} {astro-ph/9705235}\BibitemShut
  {NoStop}%
\bibitem [{\citenamefont {{Zdunik}}\ \emph {et~al.}(2006)\citenamefont
  {{Zdunik}}, \citenamefont {{Bejger}}, \citenamefont {{Haensel}},\ and\
  \citenamefont {{Gourgoulhon}}}]{2006A&A...450..747Z}%
  \BibitemOpen
  \bibfield  {author} {\bibinfo {author} {\bibfnamefont {J.~L.}\ \bibnamefont
  {{Zdunik}}}, \bibinfo {author} {\bibfnamefont {M.}~\bibnamefont {{Bejger}}},
  \bibinfo {author} {\bibfnamefont {P.}~\bibnamefont {{Haensel}}}, \ and\
  \bibinfo {author} {\bibfnamefont {E.}~\bibnamefont {{Gourgoulhon}}},\ }\href
  {\doibase 10.1051/0004-6361:20054260} {\bibfield  {journal} {\bibinfo
  {journal} {\aap}\ }\textbf {\bibinfo {volume} {450}},\ \bibinfo {pages} {747}
  (\bibinfo {year} {2006})},\ \Eprint {http://arxiv.org/abs/astro-ph/0509806}
  {astro-ph/0509806}\BibitemShut {NoStop}%
\bibitem [{\citenamefont {{Bejger}}\ \emph {et~al.}(2017)\citenamefont
  {{Bejger}}, \citenamefont {{Blaschke}}, \citenamefont {{Haensel}},
  \citenamefont {{Zdunik}},\ and\ \citenamefont
  {{Fortin}}}]{2017A&A...600A..39B}%
  \BibitemOpen
  \bibfield  {author} {\bibinfo {author} {\bibfnamefont {M.}~\bibnamefont
  {{Bejger}}}, \bibinfo {author} {\bibfnamefont {D.}~\bibnamefont
  {{Blaschke}}}, \bibinfo {author} {\bibfnamefont {P.}~\bibnamefont
  {{Haensel}}}, \bibinfo {author} {\bibfnamefont {J.~L.}\ \bibnamefont
  {{Zdunik}}}, \ and\ \bibinfo {author} {\bibfnamefont {M.}~\bibnamefont
  {{Fortin}}},\ }\href {\doibase 10.1051/0004-6361/201629580} {\bibfield
  {journal} {\bibinfo  {journal} {\aap}\ }\textbf {\bibinfo {volume} {600}},\
  \bibinfo {eid} {A39} (\bibinfo {year} {2017})},\ \Eprint
  {http://arxiv.org/abs/1608.07049} {arXiv:1608.07049 [astro-ph.HE]}
 \BibitemShut {NoStop}%
\bibitem [{\citenamefont {{Zdunik}}\ \emph {et~al.}(2008)\citenamefont
  {{Zdunik}}, \citenamefont {{Bejger}}, \citenamefont {{Haensel}},\ and\
  \citenamefont {{Gourgoulhon}}}]{2008A&A...479..515Z}%
  \BibitemOpen
  \bibfield  {author} {\bibinfo {author} {\bibfnamefont {J.~L.}\ \bibnamefont
  {{Zdunik}}}, \bibinfo {author} {\bibfnamefont {M.}~\bibnamefont {{Bejger}}},
  \bibinfo {author} {\bibfnamefont {P.}~\bibnamefont {{Haensel}}}, \ and\
  \bibinfo {author} {\bibfnamefont {E.}~\bibnamefont {{Gourgoulhon}}},\ }\href
  {\doibase 10.1051/0004-6361:20078346} {\bibfield  {journal} {\bibinfo
  {journal} {\aap}\ }\textbf {\bibinfo {volume} {479}},\ \bibinfo {pages} {515}
  (\bibinfo {year} {2008})},\ \Eprint {http://arxiv.org/abs/0707.3691}
  {arXiv:0707.3691}\BibitemShut {NoStop}%
\bibitem [{\citenamefont {{Dimmelmeier}}\ \emph {et~al.}(2009)\citenamefont
  {{Dimmelmeier}}, \citenamefont {{Bejger}}, \citenamefont {{Haensel}},\ and\
  \citenamefont {{Zdunik}}}]{2009MNRAS.396.2269D}%
  \BibitemOpen
  \bibfield  {author} {\bibinfo {author} {\bibfnamefont {H.}~\bibnamefont
  {{Dimmelmeier}}}, \bibinfo {author} {\bibfnamefont {M.}~\bibnamefont
  {{Bejger}}}, \bibinfo {author} {\bibfnamefont {P.}~\bibnamefont {{Haensel}}},
  \ and\ \bibinfo {author} {\bibfnamefont {J.~L.}\ \bibnamefont {{Zdunik}}},\
  }\href {\doibase 10.1111/j.1365-2966.2009.14891.x} {\bibfield  {journal}
  {\bibinfo  {journal} {\mnras}\ }\textbf {\bibinfo {volume} {396}},\ \bibinfo
  {pages} {2269} (\bibinfo {year} {2009})},\ \Eprint
  {http://arxiv.org/abs/0901.3819} {arXiv:0901.3819 [astro-ph.SR]}\BibitemShut
  {NoStop}%
\bibitem [{\citenamefont {{Abdikamalov}}\ \emph {et~al.}(2009)\citenamefont
  {{Abdikamalov}}, \citenamefont {{Dimmelmeier}}, \citenamefont {{Rezzolla}},\
  and\ \citenamefont {{Miller}}}]{2009MNRAS.392...52A}%
  \BibitemOpen
  \bibfield  {author} {\bibinfo {author} {\bibfnamefont {E.~B.}\ \bibnamefont
  {{Abdikamalov}}}, \bibinfo {author} {\bibfnamefont {H.}~\bibnamefont
  {{Dimmelmeier}}}, \bibinfo {author} {\bibfnamefont {L.}~\bibnamefont
  {{Rezzolla}}}, \ and\ \bibinfo {author} {\bibfnamefont {J.~C.}\ \bibnamefont
  {{Miller}}},\ }\href {\doibase 10.1111/j.1365-2966.2008.14056.x} {\bibfield
  {journal} {\bibinfo  {journal} {\mnras}\ }\textbf {\bibinfo {volume} {392}},\
  \bibinfo {pages} {52} (\bibinfo {year} {2009})},\ \Eprint
  {http://arxiv.org/abs/0806.1700} {arXiv:0806.1700}\BibitemShut {NoStop}%
\bibitem [{\citenamefont {{Fischer}}\ \emph {et~al.}(2011)\citenamefont
  {{Fischer}}, \citenamefont {{Sagert}}, \citenamefont {{Pagliara}},
  \citenamefont {{Hempel}}, \citenamefont {{Schaffner-Bielich}}, \citenamefont
  {{Rauscher}}, \citenamefont {{Thielemann}}, \citenamefont {{K{\"a}ppeli}},
  \citenamefont {{Mart{\'{\i}}nez-Pinedo}},\ and\ \citenamefont
  {{Liebend{\"o}rfer}}}]{2011ApJS..194...39F}%
  \BibitemOpen
  \bibfield  {author} {\bibinfo {author} {\bibfnamefont {T.}~\bibnamefont
  {{Fischer}}}, \bibinfo {author} {\bibfnamefont {I.}~\bibnamefont {{Sagert}}},
  \bibinfo {author} {\bibfnamefont {G.}~\bibnamefont {{Pagliara}}}, \bibinfo
  {author} {\bibfnamefont {M.}~\bibnamefont {{Hempel}}}, \bibinfo {author}
  {\bibfnamefont {J.}~\bibnamefont {{Schaffner-Bielich}}}, \bibinfo {author}
  {\bibfnamefont {T.}~\bibnamefont {{Rauscher}}}, \bibinfo {author}
  {\bibfnamefont {F.-K.}\ \bibnamefont {{Thielemann}}}, \bibinfo {author}
  {\bibfnamefont {R.}~\bibnamefont {{K{\"a}ppeli}}}, \bibinfo {author}
  {\bibfnamefont {G.}~\bibnamefont {{Mart{\'{\i}}nez-Pinedo}}}, \ and\ \bibinfo
  {author} {\bibfnamefont {M.}~\bibnamefont {{Liebend{\"o}rfer}}},\ }\href
  {\doibase 10.1088/0067-0049/194/2/39} {\bibfield  {journal} {\bibinfo
  {journal} {\apjs}\ }\textbf {\bibinfo {volume} {194}},\ \bibinfo {eid} {39}
  (\bibinfo {year} {2011})},\ \Eprint {http://arxiv.org/abs/1011.3409}
  {arXiv:1011.3409 [astro-ph.HE]}\BibitemShut {NoStop}%
\bibitem [{\citenamefont {{Heinimann}}\ \emph {et~al.}(2016)\citenamefont
  {{Heinimann}}, \citenamefont {{Hempel}},\ and\ \citenamefont
  {{Thielemann}}}]{2016PhRvD..94j3008H}%
  \BibitemOpen
  \bibfield  {author} {\bibinfo {author} {\bibfnamefont {O.}~\bibnamefont
  {{Heinimann}}}, \bibinfo {author} {\bibfnamefont {M.}~\bibnamefont
  {{Hempel}}}, \ and\ \bibinfo {author} {\bibfnamefont {F.-K.}\ \bibnamefont
  {{Thielemann}}},\ }\href {\doibase 10.1103/PhysRevD.94.103008} {\bibfield
{journal} {\bibinfo  {journal} {\prd}\ }\textbf {\bibinfo {volume} {94}},\
  \bibinfo {eid} {103008} (\bibinfo {year} {2016})},\ \Eprint
  {http://arxiv.org/abs/1608.08862} {arXiv:1608.08862 [astro-ph.SR]}
 \BibitemShut {NoStop}%
\bibitem{Sotani:2001bb} 
  H.~Sotani, K.~Tominaga, and K.~i.~Maeda,
\textcolor{blue} { Phys.\ Rev.\ D {\bf 65}, 024010 (2001)
  [gr-qc/0108060].}
\bibitem{Marranghello:2002yx} 
  G.~F.~Marranghello, C.~A.~Z.~Vasconcellos, and J.~A.~de Freitas Pacheco,
\textcolor{blue}  {Phys.\ Rev.\ D {\bf 66}, 064027 (2002)
  [astro-ph/0208456].}
\bibitem{Miniutti:2002bh} 
  G.~Miniutti, J.~A.~Pons, E.~Berti, L.~Gualtieri, and V.~Ferrari,
 \textcolor{blue}{  Mon.\ Not.\ R.\ Astron.\ Soc.\  {\bf 338}, 389 (2003)
  [astro-ph/0206142].}
\bibitem{Oechslin:2004yj} 
  R.~Oechslin, K.~Uryu, G.~S.~Poghosyan, and F.~K.~Thielemann,
\textcolor{blue}{ Mon.\ Not.\ R.\ Astron.\ Soc.\  {\bf 349}, 1469 (2004) 
  [astro-ph/0401083].}
\bibitem{Takami:2014zpa} 
  K.~Takami, L.~Rezzolla, and L.~Baiotti,
\textcolor{blue}{  Phys.\ Rev.\ Lett.\  {\bf 113}, no. 9, 091104 (2014)
  [arXiv:1403.5672 [gr-qc]].}

\bibitem{Bauswein:2011tp} 
  A.~Bauswein and H.-T.~Janka,
 \textcolor{blue}{ Phys.\ Rev.\ Lett.\  {\bf 108}, 011101 (2012)
  [arXiv:1106.1616 [astro-ph.SR]].}

\end{thebibliography}

%
\end{document}